\documentclass[article,twocolumn]{emulateapj}

\usepackage{graphicx,amssymb,amsmath,amsfonts}
\usepackage{color,subfigure}
\usepackage{natbib}
\usepackage[section]{placeins}
\usepackage{cases}
\usepackage{gensymb}

\newcommand{\appropto}{\mathrel{\vcenter{
  \offinterlineskip\halign{\hfil$##$\cr
    \propto\cr\noalign{\kern2pt}\sim\cr\noalign{\kern-2pt}}}}}

\def\app#1#2{%
  \mathrel{%
    \setbox0=\hbox{$#1\sim$}%
    \setbox2=\hbox{%
      \rlap{\hbox{$#1\propto$}}%
      \lower1.1\ht0\box0%
    }%
    \raise0.25\ht2\box2%
  }%
}

\ifdefined\kms
\renewcommand{\kms}{\,\rm km\ s^{-1}}
\else
\newcommand{\kms}{\,\rm km\ s^{-1}}
\fi

\newcommand{\ergs}{\,\rm erg\ s^{-1}}

\ifdefined\kms
\renewcommand{\mag}{\,\rm mag}
\else
\newcommand{\mag}{\,\rm mag}
\fi

\newcommand{\ev}{\,{\rm eV}}
\newcommand{\kev}{\,\rm keV}

\let\AAold\AA
\renewcommand{\AA}{\text{\AAold}}
\newcommand{\mic}{\,\mbox{$\mu$m}}
\newcommand{\cm}{\,{\rm cm}}

\newcommand{\pc}{\,{\rm pc}}
\newcommand{\kpc}{\,{\rm kpc}}

\newcommand{\ryd}{\,{\rm Ryd}}

\newcommand{\yr}{\,{\rm yr}}

\newcommand{\K}{\,{\rm K}}

\newcommand{\msun}{\,{\rm M_{\odot}}}
\newcommand{\zsun}{\,{\rm Z_{\odot}}}

\newcommand{\mG}{\,{\rm mG}}

\newcommand{\hnu}{\langle h\nu\rangle_i}

\renewcommand{\d}{{\rm d}}

\newcommand{\const}{{\rm const.}}

\DeclareRobustCommand{\ion}[2]{%
\relax\ifmmode
\ifx\testbx\f@series
{\mathbf{#1\,\mathsc{#2}}}\else
{\mathrm{#1\,\mathsc{#2}}}\fi
\else\textup{#1\,{\mdseries\textsc{#2}}}%
\fi}

\newcommand{\hi}{\text{H~{\sc i}}}
\newcommand{\hii}{\text{H~{\sc ii}}}

\newcommand{\sii}{\text{[S~{\sc ii}]}}

\newcommand{\oi}{\text{[O~{\sc i}]}}

\newcommand{\oiii}{\text{[O~{\sc iii}]}}
\newcommand{\oiiip}{\text{O~{\sc iii}}}

\newcommand{\oivp}{\text{O~{\sc iv}}}

\newcommand{\oviip}{\text{O~{\sc vii}}}
\newcommand{\oviiip}{\text{O~{\sc viii}}}

\newcommand{\fevii}{\text{[Fe~{\sc vii}]}}

\newcommand{\fexxivp}{\text{Fe~{\sc xxiv}}}

\newcommand{\neiii}{\text{[Ne~{\sc iii}]}}
\newcommand{\neiiip}{\text{Ne~{\sc iii}}}
\newcommand{\nev}{\text{[Ne~{\sc v}]}}
\newcommand{\nevp}{\text{Ne~{\sc v}}}
\newcommand{\nevi}{\text{[Ne~{\sc vi}]}}
\newcommand{\neviiip}{\text{Ne~{\sc viii}}}
\newcommand{\nexp}{\text{Ne~{\sc x}}}

\newcommand{\mgiip}{\text{Mg~{\sc ii}}}

\newcommand{\civp}{\text{C~{\sc iv}}}

\newcommand{\Ha}{\text{H$\alpha$}}

\newcommand{\Hb}{\text{H$\beta$}}

\ifdefined\aap
\else
\newcommand{\aap}{A\&A}
\newcommand{\araa}{ARA\&A}
\newcommand{\apjl}{ApJ}
\newcommand{\apjs}{ApJS}
\newcommand{\apj}{ApJ}
\newcommand{\aj}{AJ}
\newcommand{\mnras}{MNRAS}
\newcommand{\pasp}{PASP}

\newcommand{\nat}{Nature}

\newcommand{\pasj}{PASJ}
\fi
\newcommand{\rmxaa}{RevMexA\&A}

\newcommand{\nar}{NewAR}

\newcommand{\cloudy}{{\sc cloudy}}

\newcommand{\spi}{{\it Spitzer}}

\newcommand{\mbh}{M_{\rm BH}}

\newcommand{\aion}{\alpha_{\rm ion}}

\newcommand{\ebv}{E(B-V)}

\newcommand{\rsub}{r_{\rm sub}}

\newcommand{\nH}{n_{\rm H}}
\newcommand{\NH}{N_{\rm H}}

\newcommand{\pgas}{P_{\rm gas}}

\newcommand{\frad}{F_{\rm rad}}

\newcommand{\sigbar}{{\bar \sigma}}
\newcommand{\taubar}{{\bar \tau}}

\newcommand{\prad}{P_{\rm rad}}
\newcommand{\phot}{P_{\rm hot}}

\newcommand{\pmag}{P_{\rm mag}}

\newcommand{\rcrit}{r_{\rm crit}}
\newcommand{\pdot}{{\dot p}}

\newcommand{\usurf}{U(\taubar=0)}
\newcommand{\vwind}{v_{\rm s}}

\newcommand{\lbhb}{L_{\rm broad~H\beta}}
\newcommand{\lbha}{L_{\rm broad~H\alpha}}

\newcommand{\loiii}{L_{\oiii}}
\newcommand{\lx}{L_{\rm X}}

\newcommand{\loiiiHb}{\oiii/\Hb}
\newcommand{\loiiiHblong}{\oiii~5007{\rm \AA}/\Hb}
\newcommand{\ldhb}{L^{d}_{\Hb}}
\newcommand{\lir}{L_{\rm IR}}
\newcommand{\ldhbir}{\lir/\ldhb}
\newcommand{\rbreak}{r_{\rm break}}

\newcommand{\Facc}{F_{\rm acc}}
\newcommand{\Faccb}{\left< F_{\rm acc} \right>}
\newcommand{\vnuc}{v_{\rm in}}

\newcommand{\thot}{T_{\rm hot}}
\newcommand{\nhot}{n_{\rm hot}}
\newcommand{\tc}{T_{\rm C}}

\newcommand{\pratio}{\phot/\prad}
\newcommand{\pratiomean}{\left<\pratio\right>}

\begin{document}

\title{Constraining the dynamical importance of hot gas and radiation pressure in quasar outflows using emission line ratios}
\author{
Jonathan~Stern\altaffilmark{1}\footnotemark[*]\footnotemark[\textdagger], 
Claude-Andr{\'e}~Faucher-Gigu{\`e}re\altaffilmark{2},
Nadia~L.~Zakamska\altaffilmark{3} and
Joseph~F.~Hennawi\altaffilmark{1}
}
 \footnotetext[*]{E-mail: stern@mpia.de}
 \footnotetext[\textdagger]{Alexander von Humboldt Fellow}
 \altaffiltext{1}{Max Planck Institut f\"{u}r Astronomie, K\"{o}nigstuhl 17, D-69117, Heidelberg, Germany} 
 \altaffiltext{2}{Department of Physics and Astronomy and CIERA, Northwestern University, 2145 Sheridan Road, Evanston, IL 60208, USA} 
 \altaffiltext{3}{Department of Physics \& Astronomy, Johns Hopkins University, Bloomberg Center, 3400 N. Charles St., Baltimore, MD 21218, USA} 

\begin{abstract}
Quasar feedback models often predict an expanding hot gas bubble which drives a galaxy-scale outflow. 
In many circumstances this hot gas radiates inefficiently and is therefore difficult to observe directly.
We present an indirect method to detect the presence of a hot bubble using hydrostatic photoionization calculations of the cold ($\sim10^4\K$) line-emitting gas. 
We compare our calculations with observations of the broad line region, the inner face of the torus, the narrow line region (NLR), and the extended NLR, 
and thus constrain the hot gas pressure at distances $0.1\pc-10\kpc$ from the center.
We find that emission line ratios observed in the average quasar spectrum are consistent with radiation-pressure-dominated models on all scales. 
On scales $<40\pc$ a dynamically significant hot gas pressure is ruled out,
while on larger scales the hot gas pressure cannot exceed six times the local radiation pressure.
In individual quasars, $\approx25\%$ of quasars exhibit NLR ratios that are inconsistent with radiation-pressure-dominated models, 
though in these objects the hot gas pressure is also unlikely to exceed the radiation pressure by an order of magnitude or more.
The derived upper limits on the hot gas pressure imply that the instantaneous gas pressure force acting on galaxy-scale outflows 
falls short of the time-averaged force needed to explain the large momentum fluxes $\dot{p}\gg L_{\rm AGN}/c$ inferred for galaxy-scale outflows.
This apparent discrepancy can be reconciled 
if optical quasars previously experienced a buried, 
fully-obscured phase during which the hot gas bubble was more effectively confined and during which galactic wind acceleration occurred.
\end{abstract} 


\section{Introduction} 
\setcounter{footnote}{0}

Galaxy-scale outflows driven by quasars are thought to play an important role in galaxy formation, since they provide a mechanism for the central black hole (BH) to possibly regulate star formation in the host galaxy. This mechanism can potentially explain the relation between the black hole mass $\mbh$ and the galaxy bulge properties
% and the properties of the galaxy bulge 
(e.g.\ \citealt{SilkRees98, WyitheLoeb03,DiMatteo+05}), and why star formation appears to be suppressed in massive galaxies (e.g.\ \citealt{Springel+05, Hopkins+08}). 
In recent years, multiple studies have presented observational evidence for the existence of such quasar-driven galaxy-scale outflows (\citealt{Nesvadba+06,Nesvadba+11,Greene+11, RupkeVeilleux11, Sturm+11, Maiolino+12, Liu+13b, Arav+13, Arav+14, Cicone+14, Harrison+14}). 
However, the physics which govern the dynamics of these large-scale outflows is still debated.

Observed galaxy-scale outflows are commonly found to have a radial momentum outflow rate $\pdot$
in the range $1-100\, L/c$ (see compilation of observations below). Though these measurements have large uncertainty, a value of $\pdot\gg L/c$ is frequently found, regardless of the gas-phase of the outflow (ionized, neutral, molecular) and regardless of whether the measurement is based on emission lines or on absorption lines. 
Comparably large values of $\pdot$ are apparently required by theoretical models in order to suppress star formation in the host galaxy (\citealt{FaucherGiguere+12, ZubovasKing12}), and in order to explain the $\mbh-\sigma$ relation (\citealt{DeBuhr+11,DeBuhr+12,Costa+14}). 
Such large values of $\pdot$ require the wind accelerating mechanism to somehow boost $\pdot$ above the available direct radiation momentum $L/c$,  and also above the typical momentum flux found in nuclear-scale winds, which is also $\sim L/c$ (e.g.\ \citealt{Tombesi+15, Feruglio+15, Nardini+15}). 

It is not well understood how this momentum boost is achieved. 
Even if the accreting BH is covered by a thick dusty medium, radiation pressure alone is unlikely to induce a $\pdot$ which is larger than a few times $L/c$, due to instabilities which force the gas into a configuration where the infrared photons manage to escape after a few scatterings (\citealt{KrumholzThompson12, KrumholzThompson13, Novak+12, SkinnerOstriker15, TsangMilosavljevic15}, cf.~\citealt{Davis+14}). Achieving a significant momentum boost with radiation pressure is even less plausible in optical quasars, where $\sim50\%$ of sightlines are transparent even to ultraviolet radiation, and hence infrared photons will likely escape through clear paths after at most a single scattering on average. 
Alternatively, a momentum boost may not be necessary if the nuclear-scale winds have $\pdot \gg L/c$, as suggested in some models of super-Eddington accretion (\citealt{Takeuchi+13, Jiang+14, McKinney+14}), and this large momentum is transferred to the galaxy-scale outflows. However, one would then have to explain why these large-$\pdot$ nuclear winds have thus far avoided detection. 
One would also have to assume that super-Eddington accretion occurs generically in luminous quasars, an assumption for which there is at present little observational support.

A third mechanism which has been proposed to produce large momentum outflows rates was suggested by \citeauthor{FaucherGiguereQuataert12} (2012, hereafter FGQ12, see also \citealt{ZubovasKing12}). FGQ12 noted that the nuclear winds observed in UV and X-ray absorption features with velocities $\vnuc\sim10\,000\kms$ (e.g.\ \citealt{Weymann+81,Tombesi+15}) are expected to shock as they encounter the slow-moving interstellar medium (ISM). FGQ12 showed that the hot shocked gas is predicted to generically not cool efficiently (cf.\ previous work by \citealt{SilkNusser10}), and hence the hot gas bubble can do work on the surrounding ISM, thus driving a galaxy-scale outflow. In the limit of perfect energy conservation, the resulting momentum outflow rate is $\pdot \sim (\vnuc/\vwind)L/c$, where $\vwind~(\ll\vnuc)$ is the velocity of the swept-up galaxy-scale outflow. One advantage of this mechanism is the relatively small column required for a hot gas particle to transfer its momentum via collisions, compared to the 
column of $\sim10^{23}\cm^{-2}$ required to absorb an IR photon. Even so, it is still 
an open question whether the hot gas manages to drive a galaxy-scale wind with $\pdot\gg L/c$, which requires the hot shocked gas to be reasonably well confined, or whether the hot gas pressure merely vents out of the galaxy along paths of least resistance (e.g.\ \citealt{Wagner+13}). 

If accreting BHs evolve from an initial fully obscured phase into optical quasars (e.g.\ \citealt{Sanders+88, Hopkins+05}), then in principle the winds could have attained their large $\pdot$ during the fully obscured phase, when the confinement of the hot gas is more effective. In the ensuing optical quasar phase, the winds might cease to accelerate and `cruise' with their existing inertia. While these quasar `blowout' models are  plausible, they have yet to be fully established observationally.

Given the various possible scenarios discussed above, it would be valueable to detect, or rule out, the existence of hot gas bubbles in accreting BHs. A detection in optical quasars would suggest that the outflows can be accelerated by hot gas during the quasar phase. A failure to detect hot gas in quasars would suggest that either the quasar evolves and one should search for the hot gas in buried accreting BHs, or that some other mechanism is required to explain the large observed values of $\pdot$. 

The hot gas bubble cannot be detected directly by virtue of not radiating significantly. 
\cite{Nims+15} derived some indirect observational signatures of the hot bubble, 
namely the radio and X-ray emission expected when the hot gas drives a shock into the ISM of the quasar host galaxy.
Another indirect method  which can constrain the existence of a hot gas bubble was proposed by \citeauthor{YehMatzner12} (2012, hereafter YM12, see also \citealt{Yeh+13} and \citealt{Verdolini+13}), in the context of feedback in star-forming regions. YM12 demonstrated that the ionization state of the relatively cold ($T\sim10^4\K$) line-emitting gas embedded in the hot gas depends on the dominant pressure source applied to the cold gas illuminated surface. 
Specifically, YM12 used available estimates of the ionization state in the cold gas to constrain the ratio of the hot gas pressure to the incident radiation pressure. They concluded that most \hii-regions in star-forming galaxies are not significantly overpressurized by hot gas pressure, and hence cannot be the outer edges of adiabatic wind bubbles. 
In this work, we demonstrate that the YM12 method is applicable also to accreting BHs, and constrain the ratio of the hot gas pressure to the radiation pressure in a typical quasar. As we show below, the copious amount of high-energy photons emitted by quasars enables deriving strong constraints on the nature of the dominant pressure source.

To constrain the ionization state of the line-emitting gas, we utilize observed emission line ratios. 
The broad line region (BLR), with line widths of thousands of $\kms$, resides at $r \lesssim 0.1 \pc$ in $L=10^{46}\ergs$ quasars, as indicated by reverberation mapping studies (\citealt{Kaspi+05, Bentz+09}). The narrow line region, with typical line widths of hundreds of $\kms$, resides at $r\sim10-1000\pc$. Some quasars show evidence for narrow line emission from as far as $\sim10\kpc$, known as the extended NLR (e.g. \citealt{FuStockton09, Liu+13a, Greene+14}). Additionally, some quasars exhibit weaker high-ionization lines with intermediate widths (\citealt{KoristaFerland89, Rose+15}), 
and hence they likely originate from intermediate scales of $0.1\lesssim r \lesssim 10\pc$, plausibly the illuminated surface of the the IR-emitting torus (\citealt{PierVoit95, Rose+15}).
This huge dynamic range in distance of emission line regions from the nucleus enables us to probe the ratio of the hot gas pressure to the radiation pressure from nuclear scales to host-galaxy scales. 

This work builds on several previous studies, which analyzed emission line regions of active galactic nuclei (AGN) in the context of the dominant pressure source applied to the illuminated surface. \cite{PierVoit95} showed that emission line models in which radiation pressure on dust dominates the ambient gas pressure, known as radiation pressure confined (RPC) models, can explain both the wide spectrum of high-ionization lines observed in some Seyferts (\citealt{KoristaFerland89}), and the lack of significant Balmer emission from the illuminated surface of the torus.  \cite{Dopita+02} showed that RPC models can explain the uniformity of line ratios and typical ionization level in the NLR. 
\cite{Baskin+14a} showed that RPC models are also applicable to dust-less gas, where the radiation pressure is transmitted to the gas via ionization edges and resonance lines. They used dust-less RPC models of the BLR to explain both the observed small dispersion in emission line ratios, and the stratification of ionization level with distance observed in reverberation mapping studies. 
Similarly, \cite{Rozanska+06}, \cite{Stern+14b} and \cite{Baskin+14b} found that X-ray absorbing outflows (known as `Warm Absorbers') and UV-absorbing outflows (known as Broad Absorption Lines, or BALs) are also likely to be dominated by radiation pressure on ionization edges and resonance lines. 
These results were combined by \citeauthor{Stern+14a} (2014a, hereafter S14), who presented evidence that the hot gas pressure is smaller than the radiation pressure $L/(4\pi r^2 c)$ at all relevant scales, and specifically up to $r\sim10\kpc$ in luminous quasars. 
It is the goal of this paper to more rigoursly test the suggestion of S14, and hence put stricter upper limits on the hot gas pressure as a function of distance from the nucleus. 

This paper is organized as follows. 
In \S2 we present the theoretical background of the photoionization calculations we use to estimate the ratio of the hot gas pressure to the radiation pressure. 
In \S3 we compare our models with observations of quasar emission lines on scales $\sim0.1$ pc$-10$ kpc, and derive constraints on the hot gas pressure as a function of distance.  
We discuss our results in the context of models of quasar-driven winds and quasar feedback in \S4. 
We summarize our main conclusions in \S5. 

\section{Theoretical background}
\label{sec:theoretical_background}

In this section, we demonstrate that the ratio of the hot gas pressure to the radiation pressure determines the ionization state of the photoionized gas, and hence the hot gas pressure can be constrained from observations of emission lines. 
We begin with an approximate analytic derivation based on the derivations in \cite{Dopita+02} and S14, and then corroborate our analytic results with \cloudy\ numerical calculations (\citealt{Ferland+13}).

\subsection{The pressure profile of the illuminated gas cloud}\label{sec: pgas}

We consider a cloud of gas that is irradiated by the quasar. 
The distance of the cloud from the quasar is assumed to be significantly larger than the cloud size, so the incident radiation is plane-parallel. We make no further assumptions on the cloud--quasar distance, so the following analysis is equally applicable to clouds in the BLR, clouds in the NLR, and the exposed inner surface of the torus. 
We consider two types of pressure applied to the illuminated surface of the cloud. 
Pressure due to high-velocity particles, e.g.\ the thermal pressure of hot gas or the ram pressure of an incident wind, 
and the radiation pressure induced by the absorption of incident photons. 
We assume that both these pressure terms compress the gas that dominates the line emission, such that the gas pressure profile can be approximated by a hydrostatic solution. 
Note that for radiation pressure to compress the exposed gas, it must be compressed against gas which feels a weaker radiative force. For example, in a cloud that is optically thick to the ionizing radiation (`ionization-bounded' cloud), the radiation pressure can compress the ionized surface against the shielded gas beyond the ionization front. The strong BLR \mgiip\  line and NLR \sii\ and \oi\ lines observed in AGN suggest that AGN emission line clouds are typically ionization-bounded, so this requirement for the hydrostatic approximation is likely fulfilled. We further discuss the validity of the hydrostatic approximation in Appendix \ref{sec: caveats}. 

Since ram pressure and thermal pressure are transmitted to the cloud over a significantly smaller column than the column which absorbs the radiation, 
one can assume that these pressure terms set the boundary conditions at the illuminated surface, while any build up of gas pressure within the ionized layer is due to the absorption of radiation momentum. 
The boundary condition is therefore
\begin{equation}\label{eq: boundary}
\pgas(x=0) = \phot ~,
\end{equation}
where $x$ is the depth into the line-emitting cloud from the illuminated surface, $\pgas(x)$ is the thermal pressure of the line-emitting cloud, and $\phot$ represents the sum of the thermal pressure of the ambient hot gas and a possible contribution from ram pressure of an incident wind. For consistency with previous studies, we choose to use $\pgas$ to denote the thermal pressure of the $\sim10^4\K$ line-emitting gas (rather than, say, $P_{\rm cold}$). 
The hydrostatic balance equation for the gas pressure is
\begin{equation}
 \frac{\d\pgas}{\d x} = \frac{\frad(x)}{c}\nH\sigbar(x)  ~,
 \label{eq: hydrostatic}
\end{equation}
where $\frad(x)/c$ is the quasar radiation momentum flux after extinction by gas at depths smaller than $x$, $\nH$ is the hydrogen volume density, and $\sigbar$ is the spectrum-averaged extinction cross section per H-nucleon:
\begin{equation}\label{eq: sigbar definition}
 \sigbar(x)\equiv\frac{\int \sigma_\nu F_\nu(x)\d\nu }{\frad(x)} ~.
\end{equation}
Here, $F_\nu$ and $\sigma_\nu$ are the flux density and the extinction cross section at frequency $\nu$, respectively. 
The value of $\sigbar$ is set by the shape of the incident spectrum and by the composition and ionization level of the gas. 
For a typical quasar spectrum, in dusty gas $\sigbar$ will be dominated by the dust grains if the ionization parameter $U$ is $\gtrsim 0.007$, or by the \hi\ opacity at lower $U$ (\citealt{NetzerLaor93}). In dust-less gas, $\sigbar$ will be dominated by \hi\ opacity at $U\lesssim 0.05$, by metal opacity at $0.05 \lesssim U \lesssim 50$, and by electron scattering at $U\gtrsim 50$ (see, e.g.,\ figure~A1 in \citealt{Stern+14b}). 
Equation~(\ref{eq: hydrostatic}) neglects self-gravity of the absorbing cloud (a good approximation for gas that is not forming stars), radiation pressure from re-emitted radiation, and magnetic pressure, as discussed in S14 and \cite{Baskin+14a}. We further discuss the validity of neglecting magnetic pressure in Appendix \ref{sec: caveats}. 

The above definition of $\sigbar$ facilitates its use in the definition of the flux-averaged optical depth:
\begin{equation}\label{eq: taubar}
\d\taubar = \nH\sigbar \d x ~.
\end{equation}
Following \cite{Dopita+02}, we define radiation pressure as
\begin{equation}
\prad=\frac{\frad(0)}{c}= \frac{L}{4\pi r^2c} ~, 
\end{equation}
and hence we can replace $\frad(x)/c$ in eqn.~(\ref{eq: hydrostatic}) with $\prad e^{-\taubar}$. 
We emphasize that $\prad$ is defined as the momentum of the unextincted incident radiation, 
rather than the radiative momentum transferred to the gas, as used by some studies. 
The hydrostatic equation then takes the simple form 
\begin{equation}\label{eq: hydro simple form}
 \d\pgas(\taubar) = \prad e^{-\taubar}\d\taubar ~,
\end{equation}
with the solution 
\begin{equation}\label{eq: hydro solution}
 \pgas(\taubar) = \phot  + \prad\left(1-e^{-\taubar}\right) ~.
\end{equation}

Equation~(\ref{eq: hydro solution}) then implies that at $\taubar\ll 1$ 
\begin{equation}\label{eq: hydro solution tau<<1}
 \pgas(\taubar\ll 1) \sim \phot + \prad\taubar ~,
\end{equation}
while at $\taubar\gtrsim1$ 
\begin{equation}\label{eq: hydro solution tau=1}
 \pgas(\taubar\gtrsim1) \sim \phot + \prad ~.
\end{equation}
Equations (\ref{eq: hydro solution tau<<1}) and (\ref{eq: hydro solution tau=1}) show that the hydrostatic solution has two distinct regimes. If $\phot\gg\prad$, then  $\pgas$ is roughly constant and equal to $\phot$ throughout the radiation-absorbing the layer.
Such clouds are confined (on the illuminated side) by collisions-mediated forces, i.e.\ they are collisionally-confined (CC) clouds. Therefore,
\begin{equation}\label{eq: CC solution}
 P_{\rm gas,CC}(\taubar) \sim \phot ~.
\end{equation}
In the other extreme where $\phot\ll\prad$, $\pgas$ increases throughout the ionized layer, from the initial value of $\pgas=\phot$ at the illuminated surface ($\taubar=0$), to a value of $\pgas\approx\prad$ at $\taubar\gtrsim1$ near the ionization front. As noted by \cite{Dopita+02}, in such solutions $\pgas$ at $\taubar\sim1$ is independent of the boundary condition, and the cloud is Radiation Pressure Confined (RPC). 
In fact, equation~(\ref{eq: hydro solution tau<<1}) suggests that the RPC solution is independent of the boundary conditions for all layers where $\prad\taubar\gg\phot$. In this limit, RPC solutions have the simple form
\begin{eqnarray}\label{eq: RPC solution}
 P_{\rm gas,RPC}(\taubar\ll 1) &\approx& \prad\taubar \nonumber \\
 P_{\rm gas,RPC}(\taubar\gtrsim1) &\approx& \prad ~.
\end{eqnarray}

We now use this derivation of the gas pressure structure to estimate the ionization level of the gas as a function of $\phot/\prad$.

\subsection{The relation between $\prad/\pgas$ and $U$}\label{sec: Xi vs. U}

Photoionization models often employ the dimensionless ionization parameter $U$ defined as
\begin{equation}\label{eq: U}
 U \equiv \frac{\int_{\nu_0} \frac{F_\nu}{h\nu} \d \nu}{n_H c} ~,
\end{equation}
where $\nu_0$ is the hydrogen ionization frequency and the integral implicitly extends to $\nu = \infty$. 
Using $U$ is useful since it controls the ionization states of the different elements (e.g.\ \citealt{Tarter+69}), which in turn can be constrained observationally from the relative strengths of the emission lines. 
We summarize here the well-known relation between $U$ and $\prad/\pgas$ (e.g.\ \citealt{Krolik99}). Expressing $\prad$ as $\beta\int_{\nu_0} (F_\nu/c) \d\nu$, where the integral is the radiation pressure of the ionizing photons and $\beta = L/L_{\rm ion} \approx 2$ accounts for the additional pressure from the optical photons, we get
\begin{eqnarray}\label{eq: p ratio to U}
 \frac{\prad}{\pgas} &=& \frac{\beta\int_{\nu_0} F_\nu/c \d\nu}{2n_H k T} = \frac{\beta U}{2kT} \hnu \nonumber \\
 &=& 30\,U \frac{\beta}{2}\left(\frac{T}{10^4\K}\right)^{-1} \frac{\hnu}{36\ev}
\end{eqnarray}
where $k$ is the Boltzmann constant, and we use $\hnu$ to denote the average energy per ionizing photon. Note that if the gas is not dusty, then optical photons will not be absorbed in the ionized layer and the factor of $\beta$ should be dropped. 

What is the value of $U$ expected in the different scenarios discussed in the previous section?
In the collisionally confined scenario $\pgas \approx \phot$, so using equation (\ref{eq: p ratio to U}) we get
\begin{equation}\label{eq: UCC}
 U_{\rm CC} = 0.03 \frac{\prad}{\phot} \left( \frac{T}{10^4\K} \right) \left(\frac{\beta}{2}\cdot\frac{\hnu}{36\ev}\right)^{-1}~.
\end{equation}
and since $\prad\ll\phot$ we get
\begin{equation}\label{eq: col confined2}
 U_{\rm CC}\ll 0.03 \left( \frac{T}{10^4\K} \right) \left(\frac{\beta}{2}\cdot\frac{\hnu}{36\ev}\right)^{-1}~.
\end{equation}
Equation (\ref{eq: col confined2}) implies that gas confined by collisional-mediated forces must have a low ionization parameter of $U\ll 0.03$. 
Such values of $U$ are observed in LINERs, assuming that LINER-like emission originates from photoionized gas (\citealt{FerlandNetzer83, Kewley+06}). Therefore, we expect AGN with $\phot\gg\prad$ to exhibit LINER-like emission line ratios. 

On the other hand, in RPC clouds $\pgas$ increases throughout the ionized layer from $\pgas(\taubar=0)=\phot$ to $\pgas(\taubar\gtrsim1)\sim\prad$. Therefore, $U_{\rm RPC}$ will decrease from
\begin{equation}\label{eq: URPC1}
  U_{\rm RPC}(\taubar = 0) = 3\,\frac{\prad}{10\phot}\frac{T(\taubar = 0)}{10^5\K}\left(\frac{\beta}{2}\cdot\frac{\hnu}{36\ev}\right)^{-1} \\
\end{equation}
to
\begin{equation}\label{eq: URPC2}
  U_{\rm RPC}(\taubar\gtrsim1) = 0.03\, \frac{T(\taubar \gtrsim1)}{10^4\K}\left(\frac{\beta}{2}\cdot\frac{\hnu}{36\ev}\right)^{-1} ~.
\end{equation}
The gas temperature is significantly above $10^4\K$ at $\prad/\pgas\approx10$, because the metals are too ionized to be efficient coolants. 

Equation~(\ref{eq: URPC2}) suggests that most of the energy emitted from RPC clouds originates from ions that exist in gas with $U\sim 0.03$, since the layer with $\taubar\sim 1$ is the layer where most of the incident radiation energy is absorbed.
\cite{Dopita+02} showed that the values of $U$ implied by the strong optical lines in Seyferts are generally consistent with this prediction of RPC. 
The RPC scenario predicts additional (weaker) emission from the highly ionized gas which exists in the $\taubar\ll1$ surface layer seen in equation (\ref{eq: URPC1}).

Using equations~(\ref{eq: UCC})--(\ref{eq: URPC2}) one can constrain $\phot / \prad$, or equivalently $\phot / (L/4\pi r^2 c)$, from an estimate of $U$. These equations are analytical approximations intended to develop intuition. The complicated physics of line formation suggest that it is more robust to compare observed emission lines directly to predictions of detailed numerical calculations, as we do next. 

\subsection{Numerical calculations}\label{sec: numerical}

\begin{figure*}
\includegraphics{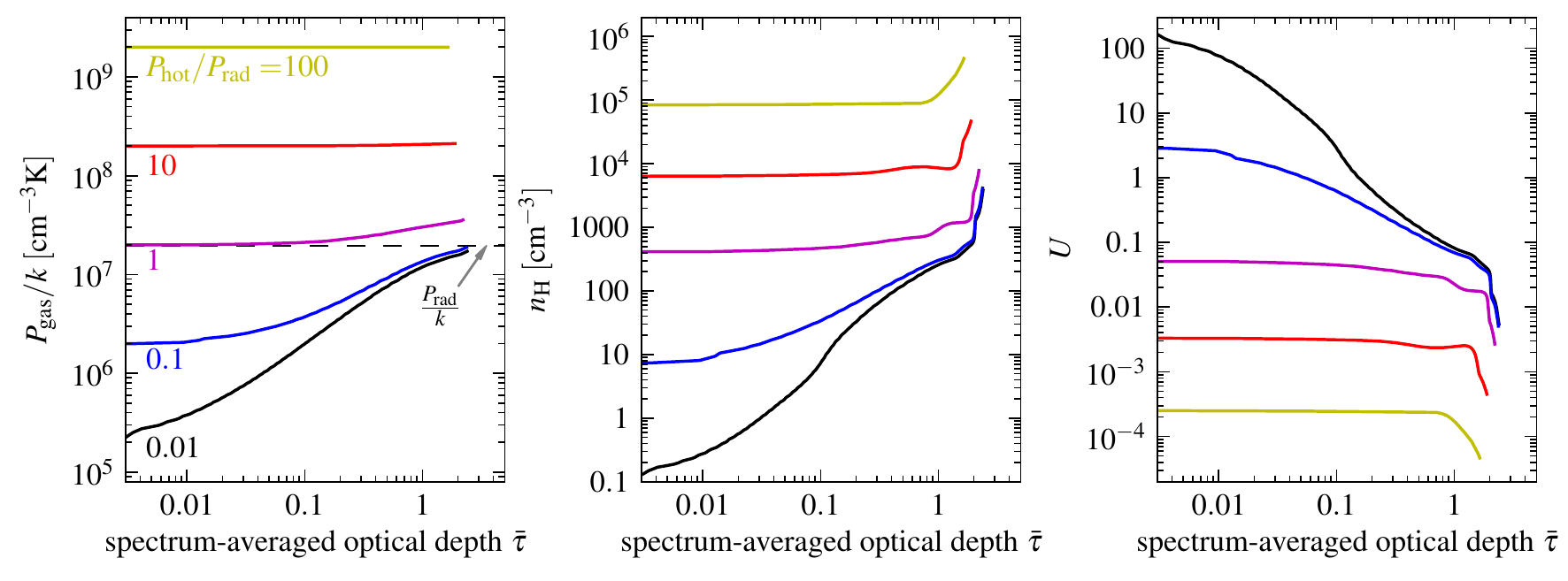}
\caption{
The structure of hydrostatic photoionized gas slabs, as a function of optical depth into the cloud and ambient hot gas pressure. 
The different colors denote models with different $\phot$, noted in the left panel relative to the assumed incident radiation pressure $\prad \equiv L/(4\pi r^2 c)$. 
All models assume a spectral slope of $\aion=-1.6$, solar metallicity, and a Milky-Way dust composition. 
The different panels plot the structure of the thermal pressure (left), gas density (middle), and ionization parameter (right), for each of the models. 
The hot gas pressure dominated models (red and yellow lines) have a near-constant value of $\pgas$ which is equal to the assumed $\phot$. These models are effectively constant density models with $U=10^{-2.5}$ or $U=10^{-3.5}$. 
In contrast, the radiation pressure dominated models (black and blue lines) show an increase of $\pgas$ throughout the slab, from the initial value of $\pgas=\phot$ at the illuminated surface ($\taubar= 0$) to $\pgas \sim \prad$ near the ionization front ($\taubar\gtrsim1$). 
The model noted by a blue line ($\phot=0.1\,\prad$) spans a range of 50 in $\nH$ at $\taubar<1$, i.e.\ in the optically thin layer of the slab. Correspondingly, the ionization parameter spans $0.06 < U < 3$ in the same range in $\taubar$. 
The model noted by a black line ($\phot=0.01\,\prad$) spans a range of 3,000 in $\nH$ at $\taubar<1$ and hence the ionization spans $0.06 < U < 200$.
}
\label{fig: hydrostatic solution}
\end{figure*}

We use version 13.03 of \cloudy\ to derive a numerical solution of hydrostatic clouds illuminated by quasar radiation. For this, we utilize the `constant pressure' flag (\citealt{Pellegrini+07}), which tells \cloudy\ to calculate $\pgas(x)$ using the assumption of hydrostatic equilibrium (eqns.~\ref{eq: boundary}--\ref{eq: hydrostatic}). We emphasize that the `constant pressure' flag tells \cloudy\ to find a hydrostatic solution, not a constant-$\pgas$ solution, as might be understood from the name.
As an instructive example, we assume a dusty ionization-bounded cloud at a distance $r=1\kpc$ from a $L_{46}=1$ quasar, where $L=10^{46}L_{46}\ergs$. The implied $\prad/k\ (= L/(4\pi r^2 c k))$ is $2\times10^7\cm^{-3}\K$. The dust grains are assumed to have a Milky-Way composition and dust-to-gas ratio. 
To explore the effect of different values of $\phot$, which is equal to the value of $\pgas$ at the cloud surface (eqn.~\ref{eq: boundary}), we assume that $\phot$ is either = $0.01$, $0.1$, $1$, $10$ or $100$ times $\prad$. We assume that the gas has solar metallicity, and that the incident spectrum is the typical AGN spectral distribution distribution (SED) seen in observations, as described in Appendix \ref{sec: spectrum}. 
The unobservable EUV part of the SED is assumed to be a simple power-law interpolation between the observed UV and X-ray, as suggested by EUV observations of high-$z$ quasars, where some of the rest-frame EUV emission is redshifted into observable wavelengths (\citealt{Laor+97, Lusso+15}). For the X-ray to optical ratio observed in $L_{46}=1$ quasars by \cite{Just+07}, the implied EUV slope is $\aion=-1.6$ ($L_\nu \propto \nu^{\aion}$). 

Figure~\ref{fig: hydrostatic solution} shows the $\pgas$, $\nH$, and $U$ structures for each value of $\phot$ relative to $\prad$, where each model is noted by a different color. The horizontal axis is the spectrum-averaged optical depth $\taubar$ (eqn.~{\ref{eq: taubar}), which we calculate from the numerical solution using $\taubar = -\ln (\frad(x)/\frad(0))$. As common in previous studies, $U$ (eqn.~{\ref{eq: U}) is calculated by dividing the ionizing photon density at the illuminated surface $F(0)/(\hnu c)$, by the value of $n_H(x)$ within the cloud. 
The calculated depth of the ionized layer is $\ll r$ in all models, justifying our assumption of a plane-parallel geometry. 

Figure~\ref{fig: hydrostatic solution} show that the collisionally-confined models with $\phot=10\,\prad$ (red) or $\phot=100\,\prad$ (yellow) have a constant $\pgas$, a constant $\nH$ at $\taubar<1.5$, and a constant $U$ equal to either $10^{-2.5}$ or $10^{-3.5}$. The independence of $\pgas$ on $\taubar$ and the values of $U$ are consistent with the analytic derivation above (eqns.~\ref{eq: CC solution} and \ref{eq: UCC}). The rise in $\nH$ at $\taubar>1.5$ is due to the drop in $T$ below $10^4\K$ beyond the ionization front, which causes $\nH$ to increase in order to maintain hydrostatic equilibrium.  
In contrast, in the RPC model with $\phot=0.1\,\prad$ (blue), $\pgas$ increases throughout the slab by a factor of $10$, from its value at $\taubar=0$ of $\pgas=\phot$ to its value at $\taubar\gtrsim1$ of $\pgas\approx\prad$. The increase in $\pgas$ is accompanied by a decrease of a factor of five in $T$ (not shown), from $T=10^5\K$ at $\taubar=0$ to $T=2\times10^4\K$ at $\taubar=1$. The combined increase in $\pgas$ and drop in $T$ implies a factor of 50 increase in $\nH$ seen in the middle panel. Correspondingly, the blue line in the right panel shows that $U$ decreases by a factor of $50$, from $U=3$ at the illuminated surface to $U=0.06$ at $\taubar=1$. The RPC model with $\phot=0.01\,\prad$ (black) has an even more extreme behavior, with $U$ spanning a range of $3\,000$, from $U=200$ at the illuminated surface to $U=0.06$ at $\taubar=1$. This behavior is consistent with the analytic derivation above (eqns.~\ref{eq: RPC solution}, \ref{eq: URPC1}, and \ref{eq: URPC2}). 
In the intermediate model with $\phot=\prad$ (magenta), the increase in $\pgas$ throughout the slab is a mild factor of two, accompanied by only a relatively mild decrease in $U$ from $0.06$ to $0.02$. 

What is the expected emission line spectrum for each value of $\phot/\prad$? 
The emission spectrum is calculated by \cloudy\ for each model, and in the next section we compare the \cloudy\ predictions with observations. 
Here, in order to develop intuition of the expected emission line spectrum as a function of $\prad/\phot$, we calculate the emission measure of different ions, defined as:
\begin{equation}\label{eq: EM}
EM_{\rm ion} \propto \int n_{\rm ion} n_e \d x ~,
\end{equation}
where $n_{\rm ion}$ and $n_e$ are the volume densities of the ion and of the electrons, repsectively. Emission measures defined in this way are rough estimates of the relative amount of cooling (i.e., line emission) from each ion.
Figure~\ref{fig: EMD} shows the emission measure distributions (EMDs) of oxygen and neon for the models plotted in Figure~\ref{fig: hydrostatic solution}, normalized by EM(\oiiip) and EM(\neiiip), respectively. We add two models with $\phot/\prad=0.3$ (cyan line) and $\phot/\prad=3$ (green line) to increase the level of detail near the $\phot=\prad$ boundary. 
As can be seen in the top panel, when $\phot/\prad$ decreases from $100$ to $3$ the oxygen EMD shifts to higher ionization states. This increase in the ionization of oxygen is expected from eqn.~(\ref{eq: UCC}), since $U$ increases with decreasing $\phot/\prad$ in the collisionally-confined regime. 
However, in the opposite RPC regime where $\phot \ll \prad$, Figure~\ref{fig: EMD} shows that the EMDs of the low ions are independent of the exact value of $\phot/\prad$. This saturation of the EMDs is expected from equation~(\ref{eq: URPC2}), which shows that $U$ is independent of $\phot/\prad$ at $\taubar\sim1$, where the low-ions are emitted.

\begin{figure}
\includegraphics{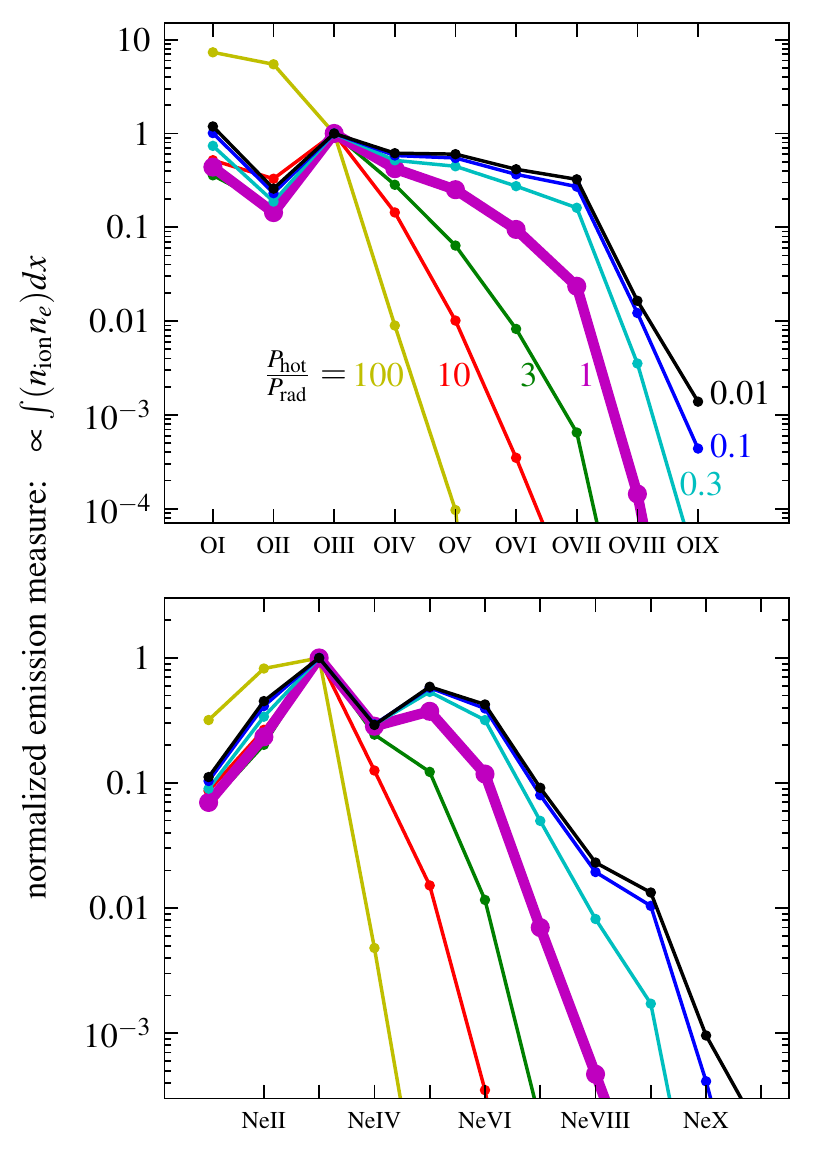}
\caption{
The emission measure distribution of oxygen and neon as a function of $\phot/\prad$. 
Models with different $\phot/\prad$ are plotted with different colors, where the $\phot=\prad$ model is emphasized with a thick line. Model parameters and line colors are as in Fig.~\ref{fig: hydrostatic solution}. The vertical axis is the emission measure (eqn.~\ref{eq: EM}), normalized by the emission measure of \oiiip\ or \neiiip, and is therefore an estimate of the expected line emission from each ion relative to the line emission from \oiiip\ or \neiiip. 
In models with $\phot/\prad\ll1$, the EMDs of the low ions are independent of $\phot/\prad$. 
Therefore, to constrain $\phot/\prad$ in this limit one needs measurements of lines from highly-ionized ions such as \oviip\ and \neviiip, which are observed at X-ray and EUV wavelengths.
}
\label{fig: EMD}
\end{figure}

Figure \ref{fig: EMD} suggests that in order to discern between the different values of $\phot/\prad$ for $\phot\ll\prad$, i.e.\ in the RPC limit, one has to observe emission lines which originate from high ions. 
For example, an observation of a \oviiip\ line is required to discern between the model with $\phot/\prad=0.3$ and models with $\phot/\prad\leq0.1$, while a $\nexp$ line is required to discern between the model with $\phot/\prad=0.1$ and models with $\phot/\prad\leq0.01$. The reason is that in order to emit high-ionization lines the slab has to have a layer with high $U$, while the maximum $U$ in an RPC slab is set by $\phot/\prad$ (eqn.~\ref{eq: URPC1}). 
As we demonstrate in the next section, focusing on the high ions is actually required even to discern between models with $\phot\sim5\,\prad$ and models with $\phot\sim\prad$, due to the uncertainty in the predicted line emission induced by the uncertainty in the other parameters of the models. 

\section{Comparison with Quasar Observations}

In this section we compare the predictions of hydrostatic photoionization models with available observations of emission lines, to constrain $\phot(r)$ at all distances from the quasar at which emission lines are observed ($0.1\pc \lesssim r \lesssim 10\kpc$). 
We focus on observations of $L\sim10^{46}\ergs$ quasars, corresponding to a $\mbh=10^9\msun$ radiating at 10\% of the Eddington limit or a $\mbh=10^8\msun$ radiating at the Eddington limit. Such luminous quasars release enough energy to potentially unbind the ISM of massive galaxies (e.g., \citealt{SilkRees98, WyitheLoeb03, King03, DiMatteo+05}), and there is growing observational evidence that only luminous AGN with $L>3\times10^{45}\ergs$ are capable of driving galaxy-wide winds (\citealt{Veilleux+13, ZakamskaGreene14}). 
Most of our analysis is focused on average quasar spectra but in \S \ref{sec:individual_objects} we examine constraints for a sample of individual objects as a way to quantify the quasar-to-quasar variance. 
The main assumptions of our photoionization models are that the gas pressure profile of the ionized layer is determined by hydrostatic balance as described in \S \ref{sec:theoretical_background} and that the emission lines originate primarily from ionization-bounded clouds. 
Other model assumptions and potential caveats are summarized in Appendix \ref{sec: caveats}.

\subsection{The extended Narrow Line Region $(r\sim1-10\kpc)$}\label{sec: eNLR}
\label{sec:eNLR}
Figure~\ref{fig: oiii to hb} compares the predictions of hydrostatic photoioinzation models to observations of kpc-scale emission in type 2 quasars. 
The horizontal axis is $\phot/\prad$, the parameter of the models that we wish to constrain. 
In order to simplify the comparison with previous studies, we note on top the ionization parameter at the illuminated surface $\usurf$. 
The value of $\usurf$ decreases with increasing $\phot/\prad$, since $\pgas=\phot$ at the surface (eqn.~\ref{eq: boundary}) and $U$ is monotonic with $\prad/\pgas$ (eqn.~\ref{eq: p ratio to U}).
The vertical axis is the predicted $\loiiiHblong$ ratio for each model. 

Each of the six plotted blue lines represents a certain combination of the gas metallicity, $Z$, and the ionizing spectral slope, which for the purpose of this study are nuisance parameters. 
We explore models with either $Z=\zsun$ or $Z=4\zsun$, corresponding to the range of gas metallicity observed in galaxies with the same mass as AGN hosts (see S14). We adopt the metal abundances as a function of overall metallicity $Z$ from \cite{Groves+04} and scale the dust mass with the metal mass. 
We also explore three possible incident spectra which differ in how we interpolate the unobservable EUV emission between the observed UV and X-ray emission. These interpolations are plotted in Appendix \ref{sec: spectrum}. 
The photoionization models assume a distance $r=1\kpc$, though the model predictions are practically identical for different $r$. 
Varying the assumed metallicity and the spectral slope enables us to ensure that the constraints derived below on $\phot/\prad$ cannot be attributed instead to uncertainties in these parameters. 
In all other aspects, the models shown in Figure \ref{fig: oiii to hb} are identical to the example models described in \S\ref{sec: numerical}. 

For each combination of metallicity and spectral slope, the predicted $\loiiiHb$ reaches an asymptotic value at $\phot/\prad\rightarrow 0$. This is the RPC limit (eqn.~\ref{eq: RPC solution}) first discussed in \cite{Dopita+02}. 
At high $\phot/\prad$ the predicted $\loiiiHb$ drops because the ionization level of the gas is too low to produce doubly ionized oxygen. 

The gray horizontal bar in Figure~\ref{fig: oiii to hb} denotes the $\loiiiHb$ ratios observed by \cite{Liu+13a}, who resolved the optical emission lines of $z\sim0.5$ type 2 quasars selected from the \cite{Reyes+08} sample. Type 2 quasars with large-scale \oiii\ emission are likely type 1 quasars viewed from an angle where the central source is obscured, rather than fully obscured accreting BHs, since in the latter case the quasar ionizing radiation could not have reached kpc scales. 	
\citeauthor{Liu+13a} measure a line ratio $\loiiiHb = 12.3\pm2.7$ constant with radius out to $r=\rbreak$, where $\rbreak = 7.0\pm1.8\kpc$. Note that the observations appear in Figure~\ref{fig: oiii to hb} as a horizontal bar since the horizontal axis is a parameter of the models, not a parameter of the observations. 
The observations are consistent with the range of line ratios allowed by RPC models, while all models with $\phot/\prad > 6$ are ruled out. 
Therefore, the observed $\loiiiHb$ at kpc scales puts an upper limit on the pressure of a putative hot gas bubble at these scales. This upper limit is equal to 
%\begin{equation}
\begin{align}
\label{eq: phot value}
\phot < 6\prad &= \frac{6L}{4\pi r^2 c} \nonumber \\
&= 1.2\cdot 10^7 L_{46}\left(\frac{r}{3\kpc}\right)^{-2}k\cm^{-3}\K.
\end{align}
%\end{equation}

\begin{figure}
\includegraphics{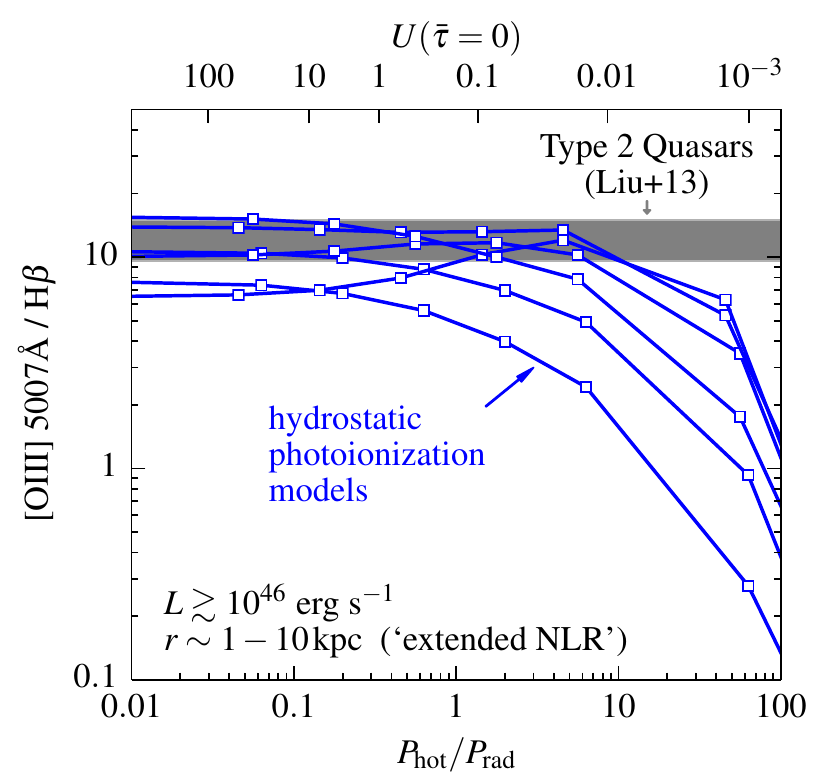}
\caption{The predictions of hydrostatic photoionization models versus observations of the extended NLR in quasars.
Blue lines denote the predicted $\loiiiHb$ as a function of the assumed $\phot/\prad$, where each line corresponds to a certain combination of gas metallicity and EUV spectral slope. 
The value of the ionization parameter at the illuminated surface appropriate for each $\phot/\prad$ is noted on top. 
The horizontal gray bar denotes the observed $\loiiiHb$ in the sample of Liu et al. (2013).
The observed $\loiiiHb$ is within the range allowed by $\phot \ll \prad$ models (RPC models), while all models with $\phot/\prad > 6$ are ruled out. 
Therefore, the observed $\loiiiHb$ at kpc scales puts an upper limit on the pressure of a putative hot gas bubble, of $\phot<6\prad\sim 10^7k\cm^{-3}\K$ (eqn.~\ref{eq: phot value}). 
}
\label{fig: oiii to hb}
\end{figure}

\begin{figure*}
\includegraphics{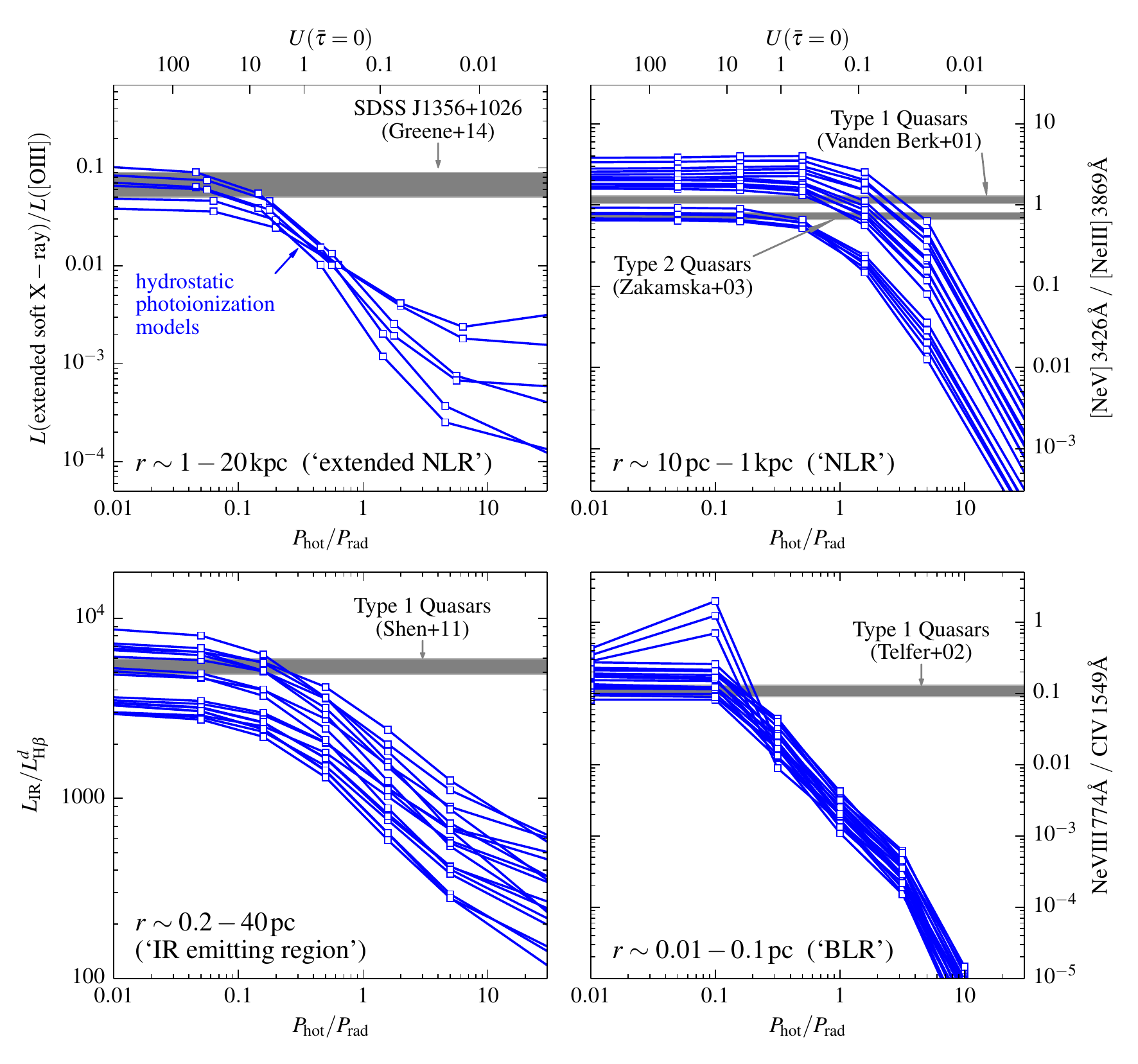}
\caption{The predictions of hydrostatic photoionization models versus emission line observations of quasars, at different distances $r$ from the nucleus. The panels are ordered by decreasing $r$, and include the eNLR (top-left), NLR (top-right), IR-emitting region (bottom-left) and BLR (bottom-right). The range of $r$ corresponding to each panel is indicated in the panel. 
Blue lines in each panel denote the hydrostatic model predictions as a function of $\phot/\prad$, where each of the 18 lines corresponds to a certain combination of the  nuisance parameters: gas metallicity, EUV spectral slope, and power-law exponent of distance distribution of the line emitting gas (eqn.~\ref{eq: eta}). The eNLR panel has only six blue lines since the radial distance $r$ is spatially resolved. The horizontal gray bars show the line ratios observed in $L_{46}\sim1$ quasars, based on mean spectra including $>$100 quasars in the NLR, IR, and BLR panels, and on a single object in the eNLR panel. 
{\bf eNLR panel:} 
the numerator of the vertical axis is the total emission at $0.5-2\kev$, which in photoionized gas is dominated by lines from highly-ionized species such as \oviip~22.1\AA\ and \oviiip~Ly$\alpha$. 
{\bf NLR panel:} 
the vertical axis is the ratio of \nev, the highest ionization line seen in the \cite{Zakamska+03} spectrum, to \neiii. 
{\bf IR panel}: the vertical axis is the ratio of the total IR emission to the \Hb-emission from dusty clouds ($\ldhb$). The observed $\lir$ is derived from the matched WISE observations of SDSS quasars, while the observed narrow component of \Hb\ is used as an upper limit on $\ldhb$. 
{\bf BLR panel}: the vertical axis is the ratio of the \neviiip\ line, the highest ionization line seen in the \cite{Telfer+02} spectrum, to the \civp\ line.
On all radial distance scales, the observations are consistent with the RPC limit ($\phot\ll\prad$), and rule out models with $\phot\gg\prad$. 
}
\label{fig: multiple line ratios}
\end{figure*}

The top-left panel of Figure~\ref{fig: multiple line ratios} is similar to Figure~\ref{fig: oiii to hb}, with $L({\rm extended~soft~X-ray})/L(\oiii)$ in the ordinate (hereafter $\lx/\loiii$). In the photoionized gas models, emission in the soft X-ray band ($0.5-2\kev$) is dominated by lines from highly ionized species such as \oviip~22.1\AA\ and \oviiip~Ly$\alpha$. The horizontal bar is the value of $\lx/\loiii$ observed in SDSS~J1356+1026 by \cite{Greene+14}. Both the Balmer ratio and the X-ray emission in this object suggest no significant extinction along the line of sight (\citealt{Greene+12, Greene+14}), so we use the directly observed value. 

The observed $\lx/\loiii$ is consistent with the range allowed by RPC models, while all models with $\phot > 0.16\,\prad$ are ruled out. That is, for $L=10^{46}\ergs$ and $r=3\kpc$ we get $\phot/k<3\times10^5 \cm^{-3}\K$. This upper limit on $\phot$ is a factor of 40 lower than the upper limit derived from $\loiiiHb$ in Fig.~\ref{fig: oiii to hb}. This difference demonstrates that emission lines from highly-ionized species enable deriving stricter upper limits on $\phot$, as implied by Fig.~\ref{fig: EMD}. 

One caveat of using the extended X-ray observation of SDSS~J1356+1026 is the lack of a high resolution X-ray spectrum, which is required to verify the photoionized nature of the emission. As discussed in \cite{Greene+14}, the soft X-ray emission is consistent with emission from photo-ionized gas, but could also originate from collisionally-ionized gas with $T\sim3\times10^6\K$. 
It is also possible that SDSS~J1356+1026 is not representative of the luminous quasar population. 
The strong constraint on $P_{\rm hot}/P_{\rm rad}$ derived from the $\lx/\loiii$ ratio assuming photoionized, ionization-bounded clouds provides a strong motivation for verifying these assumptions and pursuing other X-ray measurements. 

\subsection{The Narrow Line Region $(r\sim10\pc - 1\kpc)$}\label{sec: NLR}

We now constrain $\phot$ on scales which dominate the emission of spatially unresolved narrow lines. 
What is the distance of the unresolved NLR from the central quasar?
The minimum distance is larger than the black hole gravitational sphere of influence (otherwise the lines would not be narrow), which is $\sim 7\pc$ for a quasar with $\mbh=10^9\msun$ (eqn.~26 in S14). 
The density of the line-emitting gas must also be below the critical density of the emission line, otherwise the line emission will be collisionally suppressed. 
The relation between gas density and distance can be calculated for a given value of $\pgas/\prad$, since $\pgas/\prad \propto \nH T  L^{-1} r^2$ and $L$ and $T$ are known. The lowest values of $\pgas/\prad$ occur in the RPC limit where $\phot\ll\prad$ (left panel of Fig.~\ref{fig: hydrostatic solution}). In this limit, figure~6 in S14 shows the distances $\rcrit$ where the gas density equals the critical density for different emission lines. For the \nev\ and \neiii\ lines shown in the top-right panel of Figure~\ref{fig: multiple line ratios},  $\rcrit = 10 L_{46}^{1/2}\pc$, and line emission is collisionally suppressed at smaller $r$. For models with $\phot/\prad\gtrsim1$, the gas density is larger at a given $r$ than in the RPC limit, and therefore the gas density will equal the critical density at larger $r$. 

We use \nev\ 3426\AA\ since it is an optical line which is both strong and has relatively high ionization for an optical line. EUV or X-ray spectra are required to measure strong lines with a higher ionization level, but these are not currently available for type 2 quasars. In type 1 quasars EUV observations are available (\citealt{Telfer+02, Scott+04, Shull+12}), but the NLR component is outshined by the BLR component and quasar continuum and therefore unmeasurable. We use \neiii\ in the denominator to minimize the sensitivity of the predicted line ratio to uncertainties in abundance ratios and external dust reddening.

Models of unresolved emission have an additional free `nuisance' parameter, the distance distribution of the emission line clouds. 
This distance distribution may be important for the predicted lines ratios since different lines are collisionally de-excited at different distances, as can be seen in figure~6 of S14. 
Following S14, we parameterize the $r$-distribution as a power-law with exponent $\eta$:
\begin{equation}\label{eq: eta}
 \frac{\d \Omega}{\d (\log r)} \propto r^\eta
\end{equation}
where $\Omega$ is the covering factor of the NLR gas. The BPT line ratios (\citealt{Baldwin+81}) observed in AGN suggest that $\eta$ is likely not too far from $0$ (see Fig.~10 in S14), so we assume $\eta = -0.5, 0,$ or $0.5$. When calculating the prediction of a certain emission line, we sum the emission from models with different $r$, weighted according to equation~(\ref{eq: eta}). The integration spans the range $1\pc \leq r \leq 1\kpc$ and we compute models separated by $\d\log r=1$.
Since we calculate predictions for emission line ratios, the total $\Omega$ cancels out exactly. 
In total, we compute 18 NLR models for each value of $\phot/\prad$, corresponding to three values of $\eta$, two values of $Z$, and three spectral slopes. These models correspond to the 18 blue lines in the top-right panel of Figure~\ref{fig: multiple line ratios}. 

The relatively small dispersion observed in AGN emission line ratios (\citealt{Dopita+02}) suggests we can utilize stacked spectra to probe the typical quasar. 
We use the $\nev/\neiii$ ratios measured by \cite{VandenBerk+01} and \cite{Zakamska+03} in a stacked type 1 spectrum and a stacked type 2 spectrum, respectively. 
Narrow lines in type 2 quasars exhibit a typical extinction of $A_V=1.5\mag$ (\citealt{ZakamskaGreene14}), compared to roughly half this value in our models which include only the extinction within the dusty NLR clouds themselves, suggesting additional extinction by an unmodeled external dust screen. We correct for this unmodeled component by assuming a Small Magelanic Cloud extinction law (\citealt{WeingartnerDraine01}), which increases the observed $\nev/\neiii$ in type 2s by 15\%. 
The two observed line ratios are shown as horizontal bars in the top-right panel of Figure~\ref{fig: multiple line ratios}, where we assume a 10\% error on the measurement of the line ratio in the stacked spectra. 
Both observed $\nev/\neiii$ ratios are consistent with the range allowed by RPC models, while all models with $\prad > 3.3\,\phot\ (4.8\,\phot)$ are ruled out in type 1 (type 2) quasars. 

A similar comparison of observed narrow emission line ratios with predictions of hydrostatic models was performed by S14. The top-right panel of figure~9 in S14 shows that the observed $\nevi ~7.65\mic/\neiii~15.55\mic$ is a factor of two lower than predicted by RPC models. However, the models in S14 do not explore the range in metallicity and spectral slope explored here, but rather assume a fixed $Z=2\zsun$ and a simple power-law spectral interpolation. When allowing for these parameters to vary as done here, the observed $\nevi/\neiii$ seen in S14 is consistent with RPC models. For comparison, $\phot=10\prad$ models under-predict the observed $\nevi/\neiii$ ratio by a factor of $\gtrsim 30$, and therefore such models are robustly ruled out. 

We note that the nuclear X-ray spectrum of NGC~1068 (a lower-luminosity AGN with $L_{46} \sim 0.05$) shows emission from \fexxivp\ and other highly-ionized species (\citealt{Ogle+03}). This emission implies the existence of gas with $U\sim30$ (Fig.~9 in \citeauthor{Ogle+03}), which suggests that the NLR in such lower-luminosity AGN has $\phot/\prad < 0.1$, which is consistent with (but stronger than) the constraint derived here for quasars. 
With similar X-ray spectra, it will be possible to check whether this upper limit on $\phot$ is applicable also to luminous quasars.

\subsection{Infrared Emitting Region $(r\sim0.2-40\pc)$}\label{sec: IR}

The mean UV-selected quasar has an IR spectral energy distribution that is roughly flat in $\nu L_\nu$ in the wavelength range $3 < \lambda < 25\mic$ (\citealt{Richards+06}), and drops at longer wavelengths (\citealt{Petric+15}). The geometry of the dust which emits the near-IR and mid-IR energy is usually associated with a torodial structure known as the `torus.' 
The $\sim3\mic$ emission is expected to originate just beyond the sublimation radius ($\rsub=0.2\,L_{46}^{1/2}\pc$, \citealt{LaorDraine93}), as confirmed in luminous quasars with near-IR interferometry (\citealt{Kishimoto+11}). 
The $12\mic$ emission, according to mid-IR interferometry studies, originates at $r\approx3\pc$ in $L_{46}=1$ quasars (\citealt{Burtscher+13}). However, dust embedded in the narrow line emitting gas likely also has a significant contribution to the mid-IR emission, especially at $\lambda>15\mic$  (\citealt{NetzerLaor93, Schweitzer+08,MorNetzer12,Asmus+14}). Therefore, we adopt the radii inferred by \citeauthor{Schweitzer+08}\ of $200\,\rsub\ =40\,L_{46}^{1/2}\pc$ as the upper limit on the distance within which the IR emission is dominated in quasars.

The lower-left panel of Figure~\ref{fig: multiple line ratios} shows the hydrostatic model predictions for the ratio of $\lir$, the total IR emission, to $\ldhb$, the \Hb\ emission from the dusty gas. As in the previous section, each of the 18 blue lines corresponds to a certain combination of $Z,~\eta$ and spectral slope.
In the calculation of the predicted $\ldhb$, we include only \Hb\ emitted from the illuminated surface of the photoionized gas models; \Hb\ emission from the shielded side of the ionized layer will likely be absorbed by dust grains further in. In fact, in the limit of large column density, {\it all} the emission transmitted beyond the ionized layer will eventually be absorbed by dust grains and converted to IR radiation. Therefore, in calculating $\lir$ we include the IR emission from the illuminated surface plus all the energy transmitted past the shielded side. For a finite column density, some of the transmitted emission will not be converted to IR. Hence, the predicted $\lir$ should be treated as an upper limit. 
Also, in the surface layers of the models of the clouds with $r=1\pc$ the smallest grains are expected to sublimate. We disregard this additional complication. 

The plot shows that as $\phot/\prad$ increases from $0.1$ to $10$, $\ldhbir$ decreases by a factor of $20-30$. This behavior was explained by \cite{NetzerLaor93}, who showed that $\sigma_{\rm dust} / \sigma_{\rm gas}$ increases with $U$, where $\sigma_{\rm dust}$ and $\sigma_{\rm gas}$ are the dust and gas opacity to ionizing photons, respectively. 
Thus, in models with high $\phot/\prad$, i.e.\ a low $U$, one expects lower values of $\ldhbir$, as indeed seen in Figure~\ref{fig: multiple line ratios}. 

To derive the mean {\it observed} $\ldhbir$ in SDSS quasars, we calculate $\lir$ and $\ldhb$ based on the tables of \cite{Shen+11}. We use only quasars with $L_{46}>1$ and $z<0.8$, where \Hb\ is observable. 
Since our models predict the emission from the illuminated surface of the IR-emitting clouds, we focus on type 1 quasars where we have a clear view of the central source, and hence likely also a clear view of the illuminated surfaces of the larger scale IR-emitting clouds. 
We estimate the observed $\lir$ from $\nu L_\nu({\rm W3})$, which is the luminosity of the matched WISE (\citealt{Wright+10}) observation in the W3 band. We assume $\lir/\nu L_\nu({\rm W3})=2.7$, based on the \cite{Richards+06} mean IR SED (the k-correction is negligible since the mean IR SED is flat).
The value of $\ldhb$ is assumed to equal the narrow component of the \Hb\ line measured by \cite{Shen+11}. 
We disregard the broad component of \Hb\ since it originates from gas in the BLR, which resides within the sublimation radius, and is therefore very likely to be devoid of dust. 
Though decomposing the narrow and broad components of \Hb\ can be tricky in luminous quasars where the narrow component is weak, we find that the mean $L_{\rm narrow~\Hb} / \lbhb$ found by \cite{Shen+11} is consistent with the mean $L_{\rm narrow~\Ha} / \lbha$ found by \cite{SternLaor12b} for AGN of the same luminosity using a different fitting technique. This consistency lends credibility to the decomposition of the average line ratios used here. 
The implied mean observed $\ldhbir$ is $1.3\times10^{-4}$, which is plotted as a horizontal bar in the IR panel of Figure~\ref{fig: multiple line ratios}. 

Comparison of the predictions and observations in the IR panel of Fig.~\ref{fig: multiple line ratios} suggests that the IR-emitting gas is RPC. This result supports the suggestion of \cite{PierVoit95} that the weakness of the non-BLR Balmer emission in quasars can be explained with a RPC solution. Moreover, all models with $\phot / \prad>0.3$ are ruled out. 

In Appendix \ref{sec: caveats}, we address two potential caveats in the analysis of the IR emitting region performed here, namely extinction of \Hb, and the possibility that some of IR-emitting clouds are shielded by the BLR on smaller scales. We demonstrate there that these caveats do not significantly affect our estimate of $\phot/\prad$.

\subsection{The Broad Line Region $(r\sim0.1\pc)$}

Reverberation mapping studies suggest that the low-ionization part of the BLR resides on scales of $\approx 0.1 L_{46}^{1/2}\pc$ (\citealt{Kaspi+05}), while higher-ionization lines can originate from somewhat smaller scales.
Since these sizes are smaller than the sublimation radius, the BLR is likely devoid of dust. We therefore use \cloudy\ dust-less hydrostatic models of the BLR, similar to the models of \cite{Baskin+14a}. Due to a technical convergence problem in layers with a high $U$ in version 13.03 of \cloudy, we use version 10 of \cloudy\ (\citealt{Ferland+98}) in the BLR models, as did \citeauthor{Baskin+14a}
The models have the same range of metallicity, spectral slope and $\eta$ as the dusty models described in the previous two sections, though the integration over $r$ now spans $0.01 \leq r \leq 0.1\pc$ with $\d\log r=0.5$.  As in the dusty models above, the computed depth of the ionized layer is $<0.1\, r$ in all models, justifying our assumption of a plane-parallel geometry. 

The predictions of the hydrostatic BLR models for the $\neviiip\,\lambda774\ /\ \civp\,\lambda1549$ ratio are shown in the lower-right panel of Figure~\ref{fig: multiple line ratios}. We choose \neviiip\ since it is the highest-$U$ line observed in the BLR, and $\civp$ since it is a strong lower-ionization line which is observed simultaneously with \neviiip. As expected, $\neviiip/\civp$ increases with decreasing $\phot/\prad$ due to the increase in $U$. 

The mean observed $\neviiip/\civp$ ratio shown in Figure~\ref{fig: multiple line ratios} is taken from the mean EUV spectrum of \citeauthor{Telfer+02}\ (2002, see also \citealt{Stevans+14}), which is based on HST observations of 184 quasars. While the measured \neviiip\ luminosity may include a contribution from $\oivp~\lambda789$ (see \citeauthor{Telfer+02}), this contribution is likely subdominant since the feature is centered on 774\AA\ and is only weakly skewed to longer wavelengths. We therefore do not consider the $\oivp$ contamination as a significant caveat. Extinction by dust of the BLR typically has $\ebv\lesssim0.1\mag$ (\citealt{Richards+03, Dong+08, SternLaor12a,Krawczyk+15}), and therefore can be neglected. 
The observed $\neviiip/\civp$ is consistent with the range allowed by RPC models, while all models with $\phot/\prad>0.2$ are ruled out. Therefore, it is very likely that the BLR is RPC, as concluded by \cite{Baskin+14a}.

\subsection{Summary of the constraints on $\phot$ vs.\ $r$}

\begin{figure}
\includegraphics{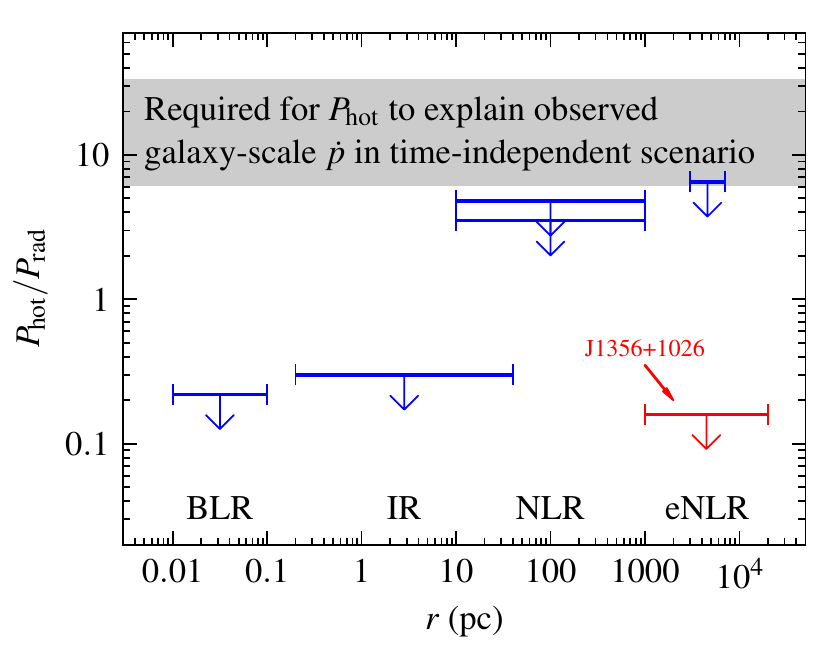}
\caption{
A summary of the derived constraints on $\phot$ as a function of the radial distance from the quasar. 
The blue upper limits are based on the comparison in Figs.~\ref{fig: oiii to hb} and \ref{fig: multiple line ratios} between hydrostatic photoionization models and emission line ratios measured in stacked quasar spectra. The red upper limit is based on the measured $\lx/\loiii$ of a single object, assuming the X-ray emission originates from photoionized gas (top-left panel in Fig.~\ref{fig: multiple line ratios}).
On all scales, the derived constraints are consistent with $\phot\ll\prad$, and rule out $\phot>6\,\prad$. 
The gray region spans the range of $\pratio$ required for the hot gas pressure to accelerate galaxy-scale outflows to their observed momentum fluxes, assuming time-steady galactic wind acceleration and quasar luminosity (eqn.~\ref{eq:pdot_lower}). This range of $\phot/\prad$ is inconsistent with the upper limits on $\phot$ implied by the emission line ratios.
}
\label{fig: Pratio vs r}
\end{figure}

Figure~\ref{fig: Pratio vs r} summarizes the upper limits derived on $\phot/\prad$ in Figures~\ref{fig: oiii to hb} and \ref{fig: multiple line ratios}. 
At all radial distances $0.01\pc \lesssim r \lesssim 10^4\pc$, the observations are consistent with the radiation pressure dominated scenario, and no other pressure source is required to explain the observations. This result is a remarkable success of the RPC mechanism. 
On scales $\sim40$ pc$-10$ kpc, optical line ratios in the average quasar spectra constrain the hot gas pressure to a modest factor of $\lesssim 6$ of the radiation pressure. A much stronger upper limit is suggested at $r\sim10$ kpc by the $\lx/\loiii$ ratio in SDSS J1356+1026, but the photoionization interpretation in this observation is less certain (see \S \ref{sec:eNLR}).

\subsection{Constraints on $\phot$ in individual objects}
\label{sec:individual_objects}
\begin{figure}
\includegraphics[trim={1cm 6cm 0 3cm},clip,scale=0.85]{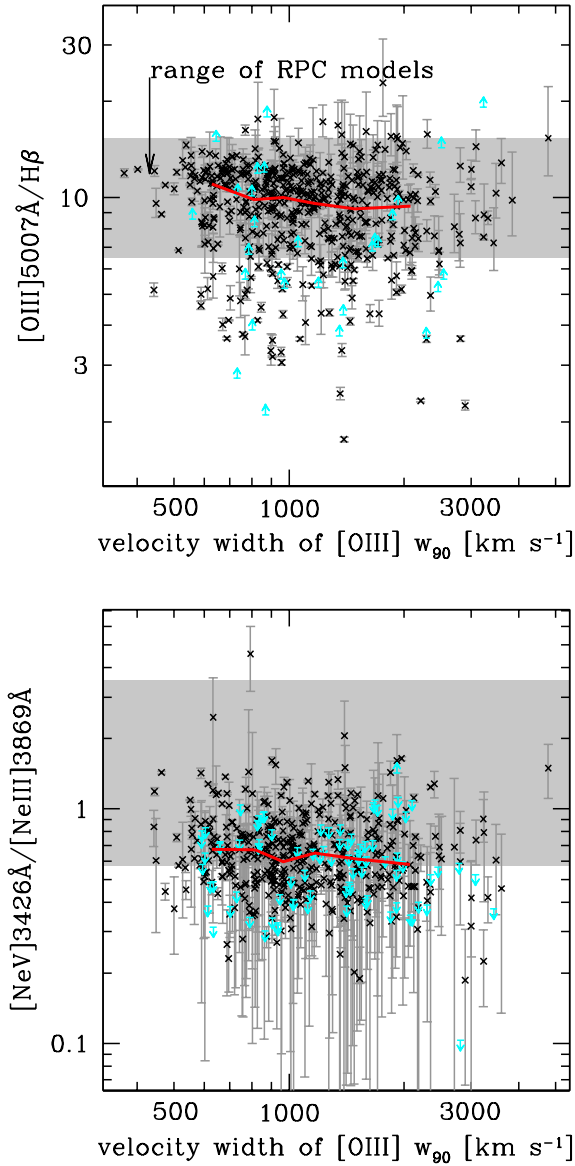} 
\caption{The dependence of line ratio diagnostics of $\phot/\prad$ on \oiii\ line kinematics, which has been suggested as a probe of outflow velocity (higher velocity outflows have higher $w_{90}$ values). Black points show objects in which both lines are confidently detected and cyan points show 3$\sigma$ lower and upper limits where only one line in the ratio is detected. Thick red lines show median values in six $w_{90}$ bins. Light grey shading shows the locus of photo-ionization models for $\phot\ll\prad$. In the top and bottom panels, 13\% and 25\% of the objects are inconsistent with RPC models, respectively. 
While the line ratios exhibit a (statistically significant) decline of $\sim$20\% when $w_{90}$ increases from 600 to $2000\kms$, this change is much weaker than expected if $\phot/\prad$ increases significantly above unity with increasing $w_{90}$, as can be seen in the NLR panel of Fig.~\ref{fig: multiple line ratios} and discussed in \S \ref{sec:individual_objects}.}
\label{fig: individual objects}
\end{figure}

The constraints on $\phot/\prad$ derived above are applicable to a typical quasar, since we compare the models with either the mean observed line ratio (Fig.~\ref{fig: oiii to hb}) or with the line ratio measured in stacked spectra (NLR, IR, and BLR panels of Fig.~\ref{fig: multiple line ratios}). 
In this section we compare the hydrostatic models predictions with line ratios in individual objects. 
Since our ultimate goal is to constrain the mechanism which drives galaxy-scale outflows, we also explore the dependence of line ratios and $\phot/\prad$ on outflow signatures. 

We use a sample of 568 optically-selected type 2 quasars with measured \oiii\ emission line kinematics (\citealt{ZakamskaGreene14}, hereafter ZG14). Specifically, ZG14 suggested that the velocity width containing 90\% of line power $w_{90}$ (measured in $\kms$) is a non-parametric measure of line kinematics which can be used as a proxy for the physical velocity of the outflow $v_{\rm out}$, with a typical scaling $w_{90}\sim (1.8-2.0) v_{\rm out}$. The optical line fluxes are measured by fitting individual Gaussians to host-subtracted spectra or by fitting the fixed-shape \oiii\ velocity profile to other emission features (\citealt{ZakamskaGreene14}); these two different measurement techniques give consistent values for the line fluxes.

In Figure~\ref{fig: individual objects}, the observed $\loiiiHb$ and $\nev/\neiii$ ratios are compared with the range of line ratios allowed by RPC hydrostatic models, which are found above to adequately explain the line ratios in the typical quasar. The gray horizontal bar now represents the predictions of the models, in contrast with Figs.~\ref{fig: oiii to hb} and \ref{fig: multiple line ratios} where the gray bar represents the observations. As discussed in \S\ref{sec: NLR}, our models do not include the effect of an external dust screen, so we reduce the predicted $\nev/\neiii$ by 15\%. Of the $\loiiiHb$ measurements, 15\% are outside the range allowed by RPC, and 12\% are inconsistent with RPC models to 1$\sigma$ of the measurement error. The respective percentages for the $\nev/\neiii$ ratio are 35\% and 25\%. This may suggest that in these objects $\phot$ is non-negligible. 
However, even in objects which are inconsistent with RPC, the typical $\nev/\neiii$ is $>0.2$, compared to the $\nev/\neiii<0.1$ expected in models with $\phot>10\prad$ (Fig.~\ref{fig: multiple line ratios}), which suggests that the hot gas pressure is unlikely to exceed the radiation pressure by an order of magnitude or more. 

The median values of the line ratios (red lines) show a weak decline with $w_{90}$, where they decrease by $\sim20\%$ from $w_{90}=600\kms$ to $w_{90}=2000\kms$. This trend is statistically significant, with 99.6\% confidence level in $\loiiiHb$ ratio and 96\% confidence level in $\nev/\neiii$. A similar trend is also seen in the mid-IR line ratio $\nevi 7.65\mic/\neiii 15.55\mic$ measured by \cite{Zakamska+16} on the subset of the sample with \spi\ observations.  However, this trend cannot reflect a significant change in $\phot/\prad$ with $w_{90}$, since $\nev/\neiii$ is expected to decrease by a much larger factor of $\sim$35 if $\phot/\prad$ increases from unity to $10$ (top-right panel of Fig.~\ref{fig: multiple line ratios}). 
Therefore, if large $w_{90}$ objects have outflowing NLRs while small $w_{90}$ objects have NLRs which are predominantly gravitationally-bound, then Figure~\ref{fig: individual objects} suggests roughly comparable $P_{\rm hot}/P_{\rm rad}$ in outflowing and non-outflowing NLRs.

\section{Discussion}\label{sec: discussion}

\subsection{Comparison with the hot gas pressure expected in simple wind models}\label{sec: expected phot}

In the previous sections, we constrained $\phot/\prad$ using observed line ratios. In this section we compare these constraints with the values of $\phot/\prad$ expected in different quasar feedback models.

\subsubsection{Compton heated wind}
First, we consider the possibility that the hot gas is fully ionized dust-less gas in thermal equilibrium with the quasar radiation, i.e.\ the hot gas temperature $\thot$ is equal to the Compton temperature $\tc\sim10^7\K$ of the quasar radiation field (e.g.\ \citealt{MathewsFerland87, Ciotti+01, Sazonov+04}). 
Figure 2 in \cite{Chakravorty+09} shows that in order for the gas to be sufficiently ionized to reach $\tc$, one must have 
\begin{equation}\label{eq: Tc}
 \left.\frac{\phot}{\prad}\right|_{\thot=\tc} < 0.05
\end{equation}
where we converted the notation of $\xi/\thot$ used by \citeauthor{Chakravorty+09}\ to the notation used here with $\xi/\thot = 8\pi c k(\prad/\phot)$. Equation~(\ref{eq: Tc}) demonstrates that in thermal equilibrium with the quasar radiation $\phot$ will always be subdominant to $\prad$. 
This result is consistent with the constraints on $\pratio$ plotted in Figure~\ref{fig: Pratio vs r}. 
However, FGQ12 show that if the hot gas bubble is driven by a fast nuclear wind, with initial velocity $v_{\rm in} \gtrsim 10,000\kms$ representative of observations of BALs, then Compton scattering is generically not efficient enough to cool the shocked wind so that thermal equilibrium with the radiation field is generally not reached. 

\subsubsection{BAL wind}\label{sec: BAL wind}

\begin{figure*}
\includegraphics[width=1.0\textwidth]{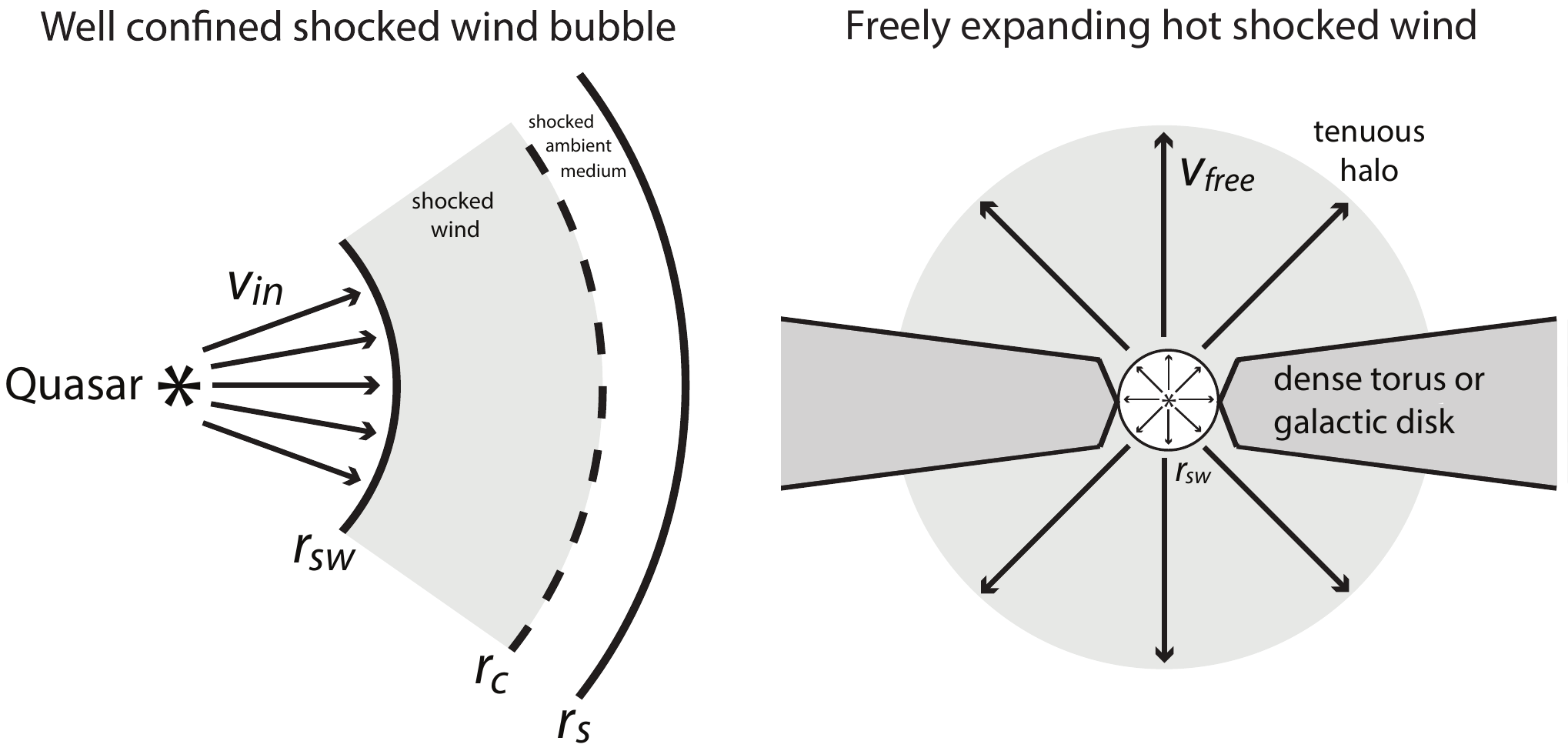}
\caption{
Two limits for the structure of the wind bubble driven by a fast (BAL-type) nuclear wind. 
\emph{Left:} The well-confined shocked wind limit, in which the hot shocked gas sweeps up considerable ambient medium mass.
\emph{Right:} The freely expanding hot shocked wind limit, in which the nuclear wind shocks when it first encounters the ISM at a small radius, but in which the hot shocked gas then expands nearly freely along paths of least resistance. 
The clouds giving rise to emission lines analyzed in this work may reside in the illuminated surface of the galactic disk or torus or be embedded in any region of the outflow.
}
\label{fig: BAL_wind_structure}
\end{figure*}

Recent models \citep[e.g.,][]{King03, Ostriker+10, FaucherGiguereQuataert12, Choi+12, Hopkins+15} indicate that the dominant form of mechanical feedback on galaxies in luminous quasars likely originates from radiatively \citep[][]{Murray+95, Proga+00} and/or magnetically accelerated \citep[][]{Emmering+92, Konigl+94} accretion disk winds. 
These winds are detected as broad absorption lines 
in the UV \citep[][]{Weymann+81, Turnshek+88} or X-ray spectra \citep[e.g.,][]{Tombesi+15, Feruglio+15, Nardini+15} of quasars. 
When such winds impact the ambient ISM, they generate a bubble of hot shocked gas, which can then sweep up gas and become mass loaded as they expand in the galaxy. 
Observations of galactic winds driven by luminous quasars on spatial scales $\sim1-10$ kpc may trace these mass loaded outflows and are inferred to have large momentum fluxes $\dot{p}\sim 1-100\,L/c$. 
We return to these observations of quasar-driven galactic winds in more detail in the next section. 

Figure \ref{fig: BAL_wind_structure} shows the expected shocked wind bubble structure in two limits: the limit of well-confined hot shocked gas sweeping up a considerable mass of ambient gas (left), 
and the limit of hot shocked gas expanding nearly freely along paths of least resistance, such as normal to the torus or galactic disk (right).
In this figure, $v_{\rm in}$ is the ``initial'' wind velocity as it leaves the accretion disk, $r_{\rm sw}$ is the radius of the (inner) wind shock, $r_{\rm s}$ is the radius of the (outer) shock with the ambient medium, and $r_{\rm c}$ is the radius of the contact discontinuity between the two shocks (see FGQ12 for more details). 
In general $r_{\rm s} \approx r_{\rm c}$ so that the swept up ambient medium accumulates at a radius $\approx r_{\rm s}$. 
We thus use $v_{\rm s}$ to denote the speed of the swept up gas (as may be observed, for example, as a kiloparsec-scale molecular outflow). 
We use $v_{\rm free}$ to denote the velocity of freely expanding hot gas. 
The clouds giving rise to emission lines analyzed in this work may reside in the illuminated surface of the torus/disk or be embedded in any region of the outflow. Next, we estimate the expected thermal pressure of the hot shocked gas in the two limits. 

{\bf Well-confined hot shocked wind:} We use the energy-conserving, spherically-symmetric wind bubble solution of FGQ12. In that solution, the thermal pressure of the shocked wind is nearly uniform with radius owing to its short sound crossing time, and it is in approximate balance with the ram pressure of nuclear wind at $r_{\rm sw}$, the radius of the wind shock:
\begin{align}
\label{eq:Phot_shock}
P_{\rm hot} \sim \frac{\dot{M}_{\rm in} v_{\rm in}}{4 \pi r_{\rm sw}^{2}},
\end{align}
where $\dot{M}_{\rm in}$ is the mass outflow rate of the small-scale wind. 
On the other hand, the radiation pressure varies with radius as the flux from the quasar drops,
\begin{align}
\label{eq:Prad}
\prad = \frac{L}{4 \pi r^{2} c},
\end{align}
so that
\begin{align}
\label{eq:Pdot over Prad well confined}
\frac{\phot}{\prad} \sim \left( \frac{\dot{M}_{\rm in} v_{\rm in}}{L/c} \right) \left( \frac{r}{r_{\rm sw}} \right)^{2}.
\end{align}
Since the natural momentum flux for a radiatively-accelerated nuclear wind is $\dot{M}_{\rm in} v_{\rm in} \sim L/c$, equation (\ref{eq:Pdot over Prad well confined}) shows that the hot gas pressure should exceed the radiation pressure in the shocked wind region, and by a large factor for $r \gg r_{\rm sw}$. 

{\bf Freely expanding hot shocked wind:} In a realistic galaxy, it is possible that the hot shocked gas expands relatively unimpeded along paths of least resistance. 
In the case of a wind driven from the nucleus at the center of a torus or galactic disk, the expansion will generally proceed normal to galactic disk, resulting in a bi-polar geometry. 
We consider the extreme limit in which the hot gas expands freely after shocking with the ambient ISM at a small radius $r_{\rm sw}$. 
As it expands, the wind cools adiabatically. 
In spherical symmetry and steady state, Appendix \ref{sec: free expansion} shows that for a monatomic gas with adiabatic index $\gamma=5/3$
\begin{eqnarray}
\label{eq:free_wind_soln}
\nhot &\propto& r^{-3/2} \\ \notag
\thot &\propto& r^{-1}
\end{eqnarray}
where $\nhot$ is the hot gas density, and hence
\begin{equation}
\phot \propto r^{-5/2}.
\end{equation}
The solution in equation (\ref{eq:free_wind_soln}) differs from the \cite[][hereafter CC85]{ChevalierClegg85} galactic wind solution ($\nhot \propto r^{-2}$, $\thot \propto r^{-4/3}$), which obeys the same equations at large $r$.
This is because the CC85 solution is assumed to be supersonic at large $r$ whereas the post-shock hot wind is subsonic in the frame of the shock. 

Using equations (\ref{eq:Phot_shock}) and (\ref{eq:Prad}) for the hot gas pressure immediately past the wind shock and the radiation pressure, we find
\begin{equation}
\frac{\phot}{\prad} \sim 
\left( \frac{\dot{M}_{\rm in} v_{\rm in}}{L/c} \right) \left( \frac{r}{r_{\rm sw}} \right)^{-1/2}
\end{equation}
so that $\phot/\prad \to 0$ as $r \to \infty$. 
This limit also holds if we assume the CC85 scalings, except that the hot gas pressure declines even faster with radius, $\phot/\prad \propto r^{-4/3}$. 
These scalings assume that the wind remains adiabatic as it expands; if the wind radiates away a substantial part of its energy on the scales of interest, then the thermal pressure of the wind gas will be even lower.

We can now use these wind models to interpret the photoionization modeling constraints on $\phot/\prad$ as a function of radius summarized in Figure \ref{fig: Pratio vs r}. 
Inside the wind shock radius $r_{\rm sw}$, the wind can be cool (as suggested by the presence of rest UV BAL features) and have negligible thermal pressure. 
This is also where the radiation pressure $\prad \propto r^{-2}$ is strongest. Therefore, for a pre-shock BAL wind we expect $\phot/\prad \ll 1$ within $r_{\rm sw}$. 
In a dynamic scenario, $r_{\rm sw}$ can move outward in time and take a wide range of values.
In a nucleus that is being evacuated by strong quasar feedback, values up to $r_{\rm sw} \sim 10-100$ pc are possible. 
Thus, the photoionization constraints of $\phot/\prad<1$ in the BLR and IR regions from $r\sim0.01\pc$ to $r\sim40\pc$ are consistent with a pre-shock BAL wind.

\begin{figure*}
\includegraphics{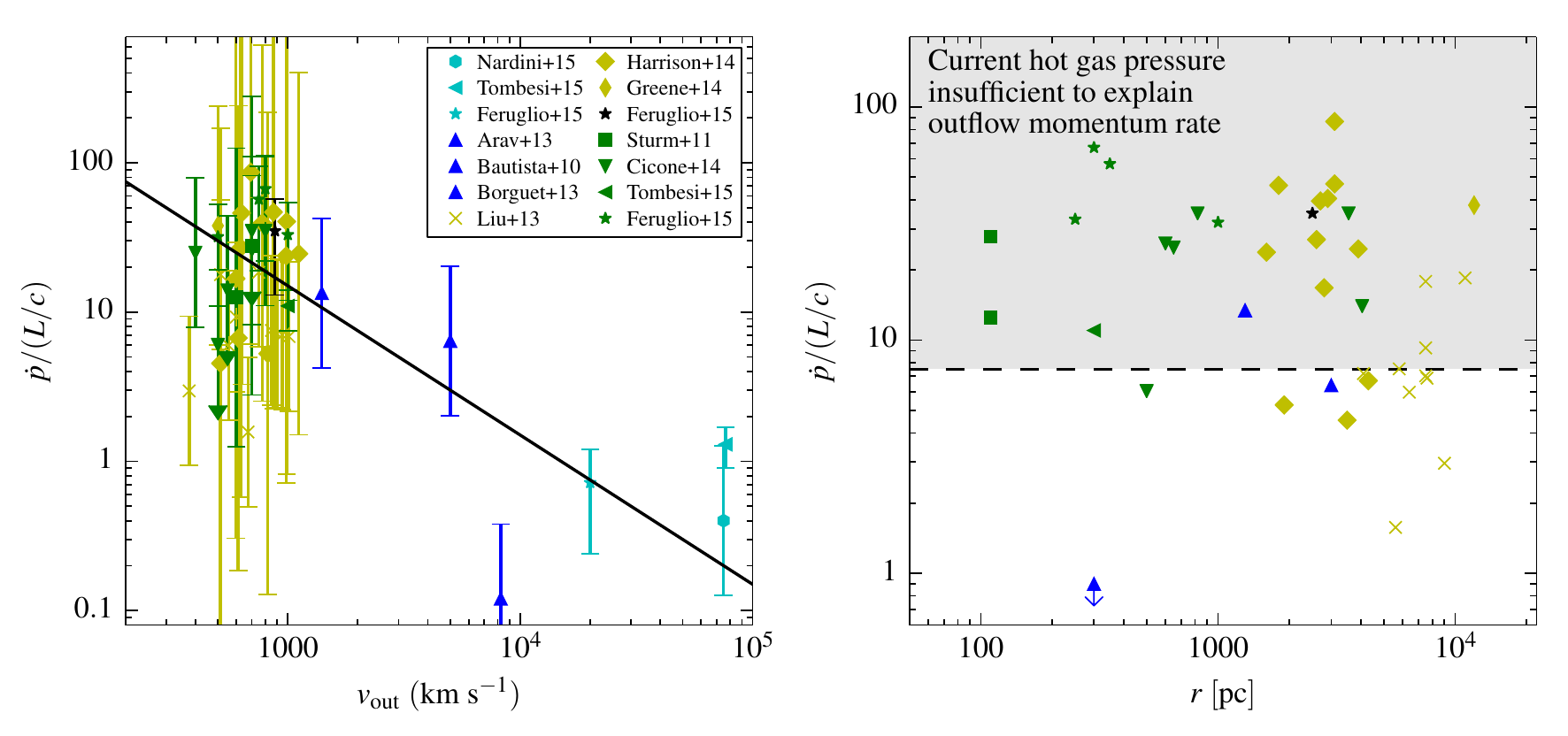}
\caption{
\emph{Left:} A compilation of outflow momentum flux measurements vs. bulk outflow velocity for galaxy-scale outflows in luminous quasars. Only quasars with luminosity $0.1<L_{46}<10$ are included. 
The reference for each observation is indicated in the legend, while the marker color denotes the gas phase in which the outflow is detected: {\it cyan}, highly ionized gas detected in X-ray; {\it blue}, ionized gas detected in UV absorption; {\it yellow}, ionized gas detected in optical emission lines; {\it black}, neutral gas; {\it green}, molecular gas. The black line show the $\pdot\propto v_{\rm out}^{-1}$ relation expected in an energy-conserving outflow, with normalization appropriate for an assumed nuclear wind speed $\vnuc=0.1c$. 
\emph{Right:} Outflow momentum flux vs. distance of the outflow from the quasar. 
Markers and colors are as in the left panel but error bars are omitted for clarity. 
The gray region shows outflow momentum fluxes that cannot be achieved by acceleration by hot gas pressure given the upper limits on $P_{\rm hot}/P_{\rm rad}$ shown in Fig.~\ref{fig: Pratio vs r}, assuming a time-steady wind and quasar luminosity. 
Most of the observed $\pdot$ are within the gray region, indicating that the current hot gas pressure is insufficient to explain the large (albeit uncertain) momentum fluxes inferred of present galaxy-scale outflows. 
This suggests that the observed galaxy-scale outflows and/or the quasar luminosity evolved systematically since the outflows were launched from the nucleus, as would be the case if the outflows obtained their large momentum fluxes in an earlier phase in which the quasars were buried and obscured in the optical (see the discussion in \S \ref{sec: discussion}). 
}
\label{fig: pdot}
\end{figure*}

At the wind shock, the wind is shock-heated to a high temperature
\begin{equation}
T_{\rm sh} \approx 1.2\times10^{10}~{\rm K} \left( \frac{v_{\rm in}}{0.1c} \right)^{2}
\end{equation}
and, in the simple wind models outlined above, the ratio of hot gas pressure to radiation pressure either increases or decreases with radius depending on whether the hot shocked gas is effectively confined or not. 
If we neglect the eNLR constraint based on $L_{\rm X} / L_{\rm [OIII]}$ (shown in red in Figure \ref{fig: Pratio vs r}), which is more uncertain since it is a based on a single object and because it is unknown whether the X-rays are powered by shocks or photoionization, then the upper limits on $\phot/\prad$ are $\approx3-6$. 
This is consistent with either no confinement of the hot gas (if $\phot/\prad \lesssim 1$) or modest confinement (if $\phot/\prad \approx 1-6$). 
Future measurements of NLR emission lines from ions with a higher ionization level than \nevp, which was used for the constraints in Figure~\ref{fig: multiple line ratios}, will be able to test more stringently the possibility of modest confinement. 
If we include the $\phot / \prad < 0.16$ upper limit from $\lx/\loiii$ at $r\sim1-20\kpc$, then the hot wind cannot be effectively confined on those scales, unless the initial wind momentum flux is $\dot{M}_{\rm in} v_{\rm in} \ll L/c$. 

\subsection{Comparison with observed quasar-driven galactic wind momentum fluxes: constraints on the time dependence of outflow acceleration}\label{sec: pdot}
The photoionization modeling of emission lines in this paper constrains the instantaneous ratio of the pressure forces acting on the emitting clouds. 
As noted in the introduction, observations of quasars suggest the presence of galaxy-scale outflows with momentum fluxes $\pdot \sim 10\,L/c$. 
The momentum fluxes of these galactic winds is defined as 
\begin{equation}
\label{eq:pdot_def}
 \pdot \approx \frac{M_{\rm s} v_{\rm s}}{t_{\rm f}} = \frac{\int \Facc \d t}{t_{\rm f}} \equiv \Faccb,
\end{equation}
where $M_{\rm s}$ is the (swept up) gas mass in the outflow, $v_{\rm s}$ is the outflow velocity, $F_{\rm acc}$ is the net force acting on the outflow, and 
\begin{equation}
 t_{\rm f} \approx r_{\rm s}/v_{\rm s} \approx 10^6 \left(\frac{r_{\rm s}}{1\kpc}\right)\left(\frac{v_{\rm s}}{1,000\kms}\right)^{-1} \yr
\end{equation}
is the flow time. 
Equation (\ref{eq:pdot_def}) shows how the momentum flux of a wind probes the time integral of the forces acting on it.

Neglecting deceleration by gravity, we can cast our constraints on $\phot$ in terms of $\Facc$:
\begin{equation}\label{eq: Facc definition}
 \Facc = 4\pi r^2\Omega_{\rm out} \left(\phot + \prad(1-e^{-\taubar_{\rm out}})\right) ~,
\end{equation}
where 
$\Omega_{\rm out}$ and $\taubar_{\rm out}$ are the covering factor and spectrum-averaged optical depth of the outflow (ignoring multiple scatterings), respectively. 
Since $\Omega \leq 1$ and $(1-e^{-\taubar_{\rm out}}) \leq 1$, and using equation (\ref{eq:Prad}) to eliminate $r^{2}$ in favor of $L$, we get
\begin{equation}\label{eq: facc to Xi}
 \frac{\Facc}{L/c} \leq 1 + \frac{\phot}{\prad}.
\end{equation}
Averaging over time, from $t=0$ (when the outflow is first launched) to $t\approx t_{\rm f}$, we thus find
\begin{equation}
\label{eq:pdot_lower}
\frac{\dot{p}}{L/c} \leq 1 + \left< \frac{P_{\rm hot}}{P_{\rm rad}} \right>.
\end{equation}
This equation assumes that $L$ did not decrease substantially with time, which is reasonable for luminous quasars radiating near their Eddington limit.

The left panel of Figure \ref{fig: pdot} shows a compilation of $\pdot$ measurements from the literature versus outflow velocity $v_{\rm out}$ (in the models discussed above, we identify $v_{\rm s}$ with $v_{\rm out}$ for galaxy-scale outflows). 
The compilation focuses on measurements in UV- or \oiii- selected quasars with $0.1<L_{46}<10$ and includes measurements for different gas phases (indicated by the color of the marker), but the selection effects of this compiled sample are not well defined.  
A few notes on individual measurements are given in Appendix \ref{sec:momentum_appendix}. 
We crudely show uncertainties on $\pdot$ of a factor of three up and down when errors are not reported by the authors. We note that the \cite{Harrison+14} measurements are not necessarily more uncertain than the other data points in this compilation in spite of the much larger quoted errors bars as these authors explored different methods of estimating $\pdot$ from ionized gas measurements. Thus, the error bars on the \cite{Harrison+14} data points are likely more representative of the true uncertainty on $\pdot$ measurements from ionized gas. 
The $\pdot$ measurements are broadly consistent with $\pdot \propto v_{\rm out}^{-1}$ (solid black line), as expected from an energy conserving flow in which the product $\pdot v_{\rm s}$ is constant (FGQ12, \citealt{ZubovasKing12}). 

Except for the three measurements based on X-ray absorption, which most likely probe the small-scale accretion disk wind \cite[e.g.,][]{Tombesi+15} and which are plotted with cyan markers, the outflows compiled in Figure~\ref{fig: pdot} are on scales $\gtrsim$100$\pc$. In the remainder of this section, we focus on these larger scale outflows, which span $0.1 \lesssim \pdot/(L/c) \lesssim 100$, with $25-75$ percentile range of $6.9<\pdot/(L/c)<35$. 
According to eqn.~(\ref{eq:pdot_lower}), this range of $\pdot$ implies a minimum $5.9< \pratiomean <34$. 
This range of $\pratiomean$ is plotted as a gray region in Figure~\ref{fig: Pratio vs r}. 
The $\pratiomean$ required by the bulk of the $\pdot$ measurements is higher than the upper limits on the present $\pratio$ implied by the emission line ratios, at all radii. 
A similar comparison is shown in the right panel of Figure~\ref{fig: pdot}. The dashed line plots the maximal $\pdot$ attainable by a wind accelerated at {\it any} $r$, in a steady-state quasar adhering to the upper-limits shown in Figure~\ref{fig: Pratio vs r}. 
The individual $\pdot$ measurements from the left panel are plotted versus their measured radial distance from the AGN (error bars omitted for clarity). Most of the winds have a momentum flux higher than the maximum value attainable by hot gas pressure acceleration in a steady-state quasar subject to the empirical $P_{\rm hot}/P_{\rm rad}$ constraints derived in this paper.

Given the large uncertainties in the $\pdot$ measurements, it is possible that the discrepancy between the instantaneous and time-averaged $\phot/\prad$ ratios indicates true values of $\pdot$ are systematically below the large estimates in Figure \ref{fig: pdot}.
However, it is interesting that measurements of outflows of different phases yield broadly consistent results regarding the high momentum fluxes of the outflows and their dependence on the speed of the galactic wind, which suggest that the momentum fluxes of the observed quasar-driven galactic winds are in fact generally well in excess of $L/c$. 

Taking the estimated outflow momentum fluxes at face value, the discrepancy between the instantaneous and time averaged $\prad/\phot$ ratio suggests that this ratio decreased over time. 
Interestingly, this result is consistent with models of quasar evolution in which accretion by the supermassive black hole is initially obscured by massive inflows into the galactic nucleus, 
until feedback from the growing black hole clears the nucleus and reveals the black hole as an optical quasar \citep[e.g.][]{SilkRees98, Fabian99, WyitheLoeb03, DiMatteo+05, Hopkins+05}. 
In such a scenario, we expect the shocked wind to be initially well confined by the obscuring medium, and reach a hot gas pressure which is $\gg\prad$ (eqn.~\ref{eq:Pdot over Prad well confined}). The large pressure on the confining shell can than accelerate an outflow with a momentum boost 
\begin{align}
\frac{\dot{p}}{L/c} & \sim \frac{1}{2} \frac{v_{\rm in}}{v_{\rm s}} \nonumber \\
& \sim 15 \left( \frac{v_{\rm in}}{0.1c} \right) \left( \frac{v_{\rm s}}{1\,000\kms} \right)^{-1}
\end{align}
from energy conservation (FGQ12, \citealt{ZubovasKing12}). 

As the outflow eventually breaks out of the nucleus and clears sight lines along which the accretion disk becomes optically visible, the hot shocked wind expands to fill the under-dense channels and flow relatively freely out of the galaxy \citep[as seen, for example, in the simulations of][]{Hopkins+15}. 
The expansion of the hot gas decreases its pressure, and in the limit of free expansion
the simple models of the previous section show that the ratio of hot gas to radiation pressure on embedded clouds drops to $<1$, 
while the momentum flux in the swept up outflowing gas observed today remains approximately constant (for outflows that are bound to the dark matter halo, the momentum will ultimately be drained out of the outflow by gravity). 
Since the quasars compiled in Figure~\ref{fig: pdot} show outflows and emission lines on scales $\gtrsim100$ pc, most of them must have cleared a large fraction of the solid angle around the black hole. 
We thus suggest that galactic winds in these luminous quasars obtained their large momentum boosts earlier in their evolution, when the quasars were buried and their shocked wind was well confined, and that the outflows are now being observed as they move outward primarily thanks to their inertia rather than instantaneous pressure forces from hot gas or radiation. 

Can we test this quasar-evolution picture using emission line ratios?
`Buried' quasars should be detectable as AGN-powered ULIRGs, which can probably be differentiated from starburst-powered ULIRGs via their bolometric luminosity (e.g.\ \citealt{Dey+08,Zakamska+16}). Optical lines emitted from the illuminated surface of such a buried quasar will likely be absorbed by the surrounding dust grains, and will not be observable. Mid-IR lines, however, are less susceptible to dust extinction, and hence may offer a glimpse into the ionization state of the hidden illuminated surface. If this surface is indeed overpressurized by a hot gas bubble, i.e.\ $\phot\gg\prad$, than we expect the lines to exhibit a low $U\ll0.03$ (eqn.~\ref{eq: col confined2}). As noted above such low $U$ are typical of LINERs. Therefore, a quasar in the early stage of its evolution, during which the outflow is being accelerated, should appear as a ULIRG with LINER-like MIR ratios. The expected MIR line ratios of LINERs has been calculated by \cite{SpinoglioMalkan92}.

Another variant of the time evolution scenario is one in which the radiative luminosities of the quasars analyzed in this work have systematically dropped since the winds were launched from the nucleus, so that $\pdot / (L/c)$ measured at the present time would overestimate the momentum boost relative to $L/c$ at the time the outflow was launched from the nucleus. While we cannot rule out this possibility, the fact that $\pdot / (L/c) \gg L/c$ for nearly all galaxy-scale outflows in luminous quasars would require that the quasars are almost never more intrinsically luminous today than when the outflows were launched. Because we focus on luminous quasars that cannot be radiating too far from their Eddington limit, it also seems unlikely that this effect alone could explain a systematic momentum boost of more than a factor of a few. 

\subsection{Other possible wind-acceleration scenarios: radiation pressure on dust grains and super-Eddington accretion flows}
In the above discussion, we focused on the case of a small-scale (BAL) wind accelerated from the accretion disk of a supermassive black hole with a momentum flux $\lesssim L/c$ and which then powers an energy-conserving wind bubble whose momentum is enhanced during its expansion. 
Two other scenarios could also explain observations of quasar-driven galactic winds with $\dot{p} \gg L/c$. 

The first is the case of an outflow accelerated by radiation pressure on dust grains \citep[e.g.,][]{Murray+05, Roth+12, Thompson+15}. 
However, momentum fluxes well in excess of $L/c$ can only be reached in this scenario if reprocessed IR photons scatter multiple times with the outflow, i.e.\ if reprocessed radiation is effectively trapped by the outflow. 
This can certainly not be the case at the present time in the optically visible quasars whose line emitting gas we have modeled in this work since we directly see the accretion disk in optical or UV light. 
In an earlier evolutionary phase in which the quasars were buried IR photons could have been more effectively trapped. 
Thus if multiple scatterings of IR photons explains the large momentum fluxes of quasar outflows, it requires a time-dependent scenario as discussed above for the energy-conserving model. 

Whether the IR photons can provide enough extra momentum depends on whether the IR photons can avoid rapidly leaking out of the medium through the action of the radiation-hydrodynamic Rayleigh-Taylor instability. 
Several recent radiation-hydrodynamic simulations in the context of AGN and star formation suggest that when interactions between matter and radiation are self-consistently taken into account IR photons typically do not provide a momentum boost of more than a factor of a few \citep[e.g.,][]{Novak+12, KrumholzThompson13, SkinnerOstriker15}. 
This result is however sensitive to the numerical method employed (e.g.,\ \citealt{Davis+14,TsangMilosavljevic15}) and a self-consistent calculation with initial and boundary conditions representative of a quasar nucleus has not yet been performed, so it is not yet clear whether powerful outflows can be accelerated by radiation pressure on dust. 
Unlike the energy-conserving model, radiation pressure on dust grains does not naturally suggest the $\dot{p} \propto v_{\rm s}^{-1}$ trend suggested by Figure~\ref{fig: pdot}. 

A final possibility is that luminous quasars grow at super-Eddington rates in radiatively inefficient episodes. 
Radiation-magnetohydrodynamic simulations of accretion flows around black holes have demonstrated that accretion at super-Eddington rates is indeed possible if sufficient matter is supplied to the BH \citep[e.g.,][]{Takeuchi+13, Jiang+14, McKinney+14}. 
In that case the wind can have a momentum flux in excess of $L/c$ as it leaves the accretion disk and would not need to be boosted by a large factor to explain observed galaxy-scale outflows. 
However, this scenario would require that most luminous quasars are accreting super-critically, for which there is presently no observational support. 

\section{Conclusions}
In this paper, we use observations of quasar emission lines and hydrostatic photoionization models to constrain the pressure of the intercloud hot gas $\phot(r)$ relative to radiation pressure on scales $0.1\pc \lesssim r \lesssim 10\kpc$, where $r$ is the distance from the quasar. 
Our focus is on luminous quasars in which powerful galaxy-scale outflows have recently been observed in ionized, atomic, and molecular gas. 
Our main conclusions can be summarized as follows: 

\begin{enumerate}
 
\item The mean emission line ratios observed in UV-selected type 1 quasars and \oiii-selected type 2 quasars with $L\sim 10^{46}\ergs$ are consistent with $\phot(r) \ll L/(4\pi r^2 c)$ on all scales, as suggested by \cite{Stern+14a}. 
This result implies that Radiation Pressure Confined (RPC) models, proposed by \cite{Dopita+02} for the NLR, by \cite{PierVoit95} for the Torus, and by \cite{Baskin+14a} for the BLR, correctly predict a wide range of emission line ratios in the mean quasar spectrum. 

\item A dynamically important hot gas pressure is however not ruled out on scales $\gtrsim 40\pc$. On these scales optical line ratios constrain the hot gas pressure to a modest factor of $\lesssim 6$ of the radiation pressure for an average quasar spectrum. If the measured $\lx/\loiii$ ratio measured for the J1356+1026 quasar is representative of the underlying quasar population, and if the soft X-ray emission is confirmed to be of photoionization origin, then $\phot/\prad$ would be tightly constrained to $\lesssim 0.1$ in the extended NLR ($r\sim10$ kpc). This highlights the powerful diagnostic power of X-ray observations and provides a strong motivation for further X-ray analyses.

\item In individual quasars, $\approx$25\% of the objects exhibit narrow line ratios which are inconsistent with RPC models by a factor of $\sim 2$. Even in these objects, though, the hot gas pressure acting on the line-emitting clouds is unlikely to exceed the radiation pressure by an order of magnitude or more. 

\item The upper limits on $\phot$ found in this study imply that the instantaneous accelerating force due to hot gas pressure on galaxy-scale outflows cannot exceed $7L/c$ on any scale. 
For comparison, previous studies found galaxy-scale outflows in quasars with momentum outflow rates $\pdot$ of $7-35$ times $L/c$, though with large uncertainties. Taking the $\pdot$ measurements at face value, they imply a time-averaged force of $7-35$ times $L/c$. This apparent discrepancy between the time-averaged force and the instantaneous force can be reconciled if the force decreases with time. This picture is consistent with models of quasar evolution in which accretion is initially fully-obscured and hot gas resulting from the shocked accretion disk wind is well confined by the ambient medium, until feedback clears the nucleus and reveals the black hole as an optical quasar. Once the quasar becomes optically visible, the hot shocked gas may become relatively free to expand out of the galaxy along clear sight lines and its pressure drops. This scenario can be tested using mid-IR emission line ratios in fully-obscured quasars. 

\end{enumerate}

\section{Acknowledgements}
JS acknowledges financial support from the Alexander von Humboldt foundation. CAFG is supported by NSF through grants AST-1412836 and AST-1517491, by NASA through grant NNX15AB22G, and by Northwestern University funds. 
We thank the organizers of the conference `Powerful AGN and Their Host Galaxies Across Cosmic Time' in Port Douglas, Australia, for facilitating the discussion which led to this manuscript. 
CAFG acknowledges useful discussions with Eliot Quataert, Phil Hopkins and Norm Murray on the physics of AGN-driven galactic winds.

\bibliographystyle{apj}
% \nocite{*}

\appendix

\section{A. The incident quasar spectrum}\label{sec: spectrum}
Figure~\ref{fig: slopes} plots the three possible quasar spectra that we explore in this work. The optical spectral index of $-0.5$ is based on \cite{VandenBerk+01}, while we assume an X-ray spectral index of $-1$ at $2-200\kev$ (\citealt{Tueller+08, Molina+09}) and a cutoff at higher frequencies. 
Spectral indices $\alpha$ are defined as $L_{\nu}\propto \nu^{\alpha}$. 
The ratio of optical to X-ray flux is set by the mean ratio observed by \cite{Just+07} for the $L_{46}\sim1$ quasars analyzed in this work. The interpolation at unobservable EUV wavelengths is significantly more uncertain. Our fiducial model (used in Figs.~\ref{fig: hydrostatic solution} and \ref{fig: EMD}) is a single power-law interpolation between the observed UV and X-ray, as suggested by the $1.5-2\ryd$ emission observed in high-redshift quasars (\citealt{Telfer+02, Shull+12, Lusso+15}) and by the $0.2\kev~(15\ryd)$ emission observed in low-redshift quasars (\citealt{Laor+97}). 
To estimate the uncertainty in the predicted line ratios due to the uncertainty in the EUV, we consider the two additional quasar spectra shown in Figure \ref{fig: slopes} for the models shown in Figures~\ref{fig: oiii to hb} and \ref{fig: multiple line ratios}. We interpolate between the observed UV and X-ray using a broken power-law, varying the $4\ryd$ flux by a factor of ten higher or lower than in the fiducial spectrum.

\begin{figure}
\begin{center}
\includegraphics[width=0.65\textwidth]{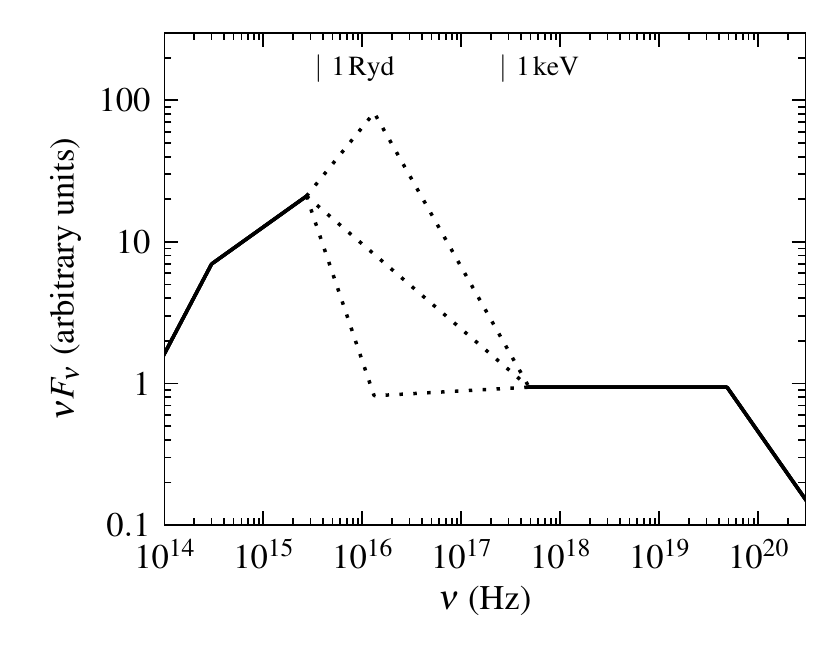}
\end{center}
\caption{The three possible quasars spectral energy distributions (SEDs) considered in this study. The solid line denotes the observable part of the spectrum. The middle dashed line is a power-law interpolation between the UV and X-ray, while the other two spectra have a $4\ryd$ flux which is a factor of ten higher or lower than in the single power-law interpolation. We estimate the uncertainty in the predicted line ratios due to the uncertainty in the $1\ryd - 2\kev$ quasar emission by exploring the effects of these three SEDs on our photoionization calculations.}
\label{fig: slopes}
\end{figure}

\section{B. Potential Caveats in the Photoionization Modeling}\label{sec: caveats}
This section discusses potential caveats in the photoionization modeling performed in this paper, and how they may affect our derived upper limits on $\phot/P_{\rm rad}$. 

\subsection{The hydrostatic gas pressure profile assumption}\label{sec: hydrostatic assumption}

In this study, the gas-pressure structure of the line-emitting gas is assumed to be determined by hydrostatic equilibrium, with either a hot gas bubble (in the $\phot\gg\prad$ scenario), or with the radiation pressure from the quasar (in the $\prad\gg\phot$ scenario). Is this assumption justified? 
In the $\prad\gg\phot$ limit, the hydrostatic assumption implies that the radiation pressure compresses the illuminated surface layer against the large inertia of the cloud. Is this configuration stable? At the illuminated side of the cloud, the accelerating force of the radiation has the same direction as the density gradient (shown in the middle panel of Fig.~\ref{fig: hydrostatic solution}), and therefore the ionized surface is Rayleigh-Taylor stable. 
However, a radiation pressure dominated cloud will be subject to ablation and evaporation from it lateral boundaries. 
The timescale of these cloud destruction mechanisms is of the order of the sound crossing time along the direction perpendicular to the radiation. 
For comparison, the time required to build the hydrostatic gas pressure profile is the sound crossing time of the ionized layer in the radial direction. In an ionization-bounded cloud where the ionized layer is only a thin surface of a much larger cloud, the former timescale is much longer than the latter. We therefore expect the ionized layer to be in hydrostatic equilibrium with the radiation pressure throughout most of the lifetime of the clouds. 
This conclusion is supported by the radiation-hydrodynamic simulation ran by \cite{Namekata+14}, which follows the evolution of an initially uniform-density spherical cloud exposed to AGN radiation. Their `High-$U$' simulation, which assumes $\prad\gg\phot$, shows the formation of a density profile at the illuminated surface in which the density rises exponentially with depth (fig.~12c in \citealt{Namekata+14}). This density profile is predicted by the hydrostatic approximation (eqn.~14 in S14), consistent with our assumption that the ionized surface is in quasi-hydrostatic equilibrium with the incident radiation. 

The radiation pressure of the IR photons re-emitted by dust grains, which may cause instabilities in large column density clouds (e.g., \citealt{KrumholzThompson13, Davis+14}), is unlikely to have a significant effect on the structure of the ionized layers modeled in this study. This follows from the small optical depth of the ionized layer to IR photons, which is $\sim \sigma_{IR} / \sigma_{UV} \sim 0.01 - 0.1$, where $\sigma_{\rm IR}$ and $\sigma_{\rm UV}$ are the dust opacities to IR and UV photons, respectively. 
Another potential concern is that the line-emitting wind clouds may be subject to destruction by the usual Kelvin-Helmholtz and Rayleigh-Taylor hydrodynamical instabilities \citep[e.g.,][]{Klein+94, Cooper+09}. These instabilities probably generally do not prevent the line-emitting clouds from reaching hydrostatic balance since the cloud sound crossing times are smaller than the time scale for the instabilities to destroy the clouds in a broad range of circumstances \citep[e.g.,][]{FaucherGiguere+12, Zhang+15}, though the physics of cloud destruction by winds and shocks is sensitive to factors such as geometry, cooling, thermal conduction, and magnetic fields \citep[e.g.,][]{Orlando+15, McCourt+15, ScannapiecoBruggen15} so that it would be useful to study this problem in more detail using numerical simulations tuned to the conditions of the line-emitting clouds in quasar-driven outflows.

\subsection{Matter-bounded clouds}
Ionization-bounded clouds have a column density $\NH$ which is larger than the column of the ionized layer. Matter-bounded clouds, in contrast, have a $\NH$ which is smaller than the potential column of the ionized layer and the entire cloud is ionized. All models used above are ionization-bounded, i.e.\ the \cloudy\ calculation continues until the \hii\ fraction drops below 10\%. 
This assumption is supported observationally, since the BLR exhibits a strong \mgiip\ line and the NLR exhibits strong \sii\ and \oi\ lines. These lines have an ionization potential of $<1\ryd$, and therefore originate from beyond the \hii\ ionization front (e.g.\ \citealt{OsterbrockFerland06}), which exists only if the clouds are ionization-bounded. 

Would including matter-bounded clouds in the modeling change the limits on $\phot$ derived in this paper?
To derive an estimate, we use the model with $\phot/\prad=10$ shown in Figures~\ref{fig: hydrostatic solution} and \ref{fig: EMD}, and assume a matter-bounded cloud with $\NH=2\times10^{20}\cm^{-2}$ instead of the $\NH>10^{21}\cm^{-2}$ assumed in the ionization-bounded case. The predicted $\nev/\neiii$ ratio shown in the top-right panel of Figure~\ref{fig: multiple line ratios} increases by a modest factor of $\sim 2$ (estimated using the emission measure, eqn.~\ref{eq: EM}) compared to the ionization-bounded scenario. This increase in the expected line ratio will increase the implied upper limit on $\phot$ for the NLR of type 2 quasars from $4.8\,\prad$ to $\sim6.3\,\prad$ in the extreme case that {\it all} NLR clouds have $\NH=2\times10^{20}\cm^{-2}$. A smaller value of $\NH$ is unlikely, since the \neiii\ line emission per unit covering factor decreases with $\NH$ and for $\NH=2\times10^{20}\cm^{-2}$ clouds the emission is already a factor of $\sim6$ smaller than in the ionization-bounded case. This would 
imply a NLR covering factor of 60\%, i.e. $\sim6$ times larger than derived assuming ionization-bounded clouds (e.g.\ \citealt{BaskinLaor05}). Therefore, the assumption of ionization-bounded clouds appears to be a good one for our purposes.

The weak dependence of predicted line ratios on assumed column applies only for hot gas pressure dominated models, in which the ionization parameter $U$ is independent of the depth into the cloud (right panel of Fig.~\ref{fig: hydrostatic solution}). In RPC clouds, the decrease of $U$ with $\taubar$ implies that high-ion and low-ion emission originate from different layers within the cloud, and therefore matter-bounded clouds would predict significantly larger high-ion to low-ion line ratios than predicted above assuming ionization-bounded clouds. Since this effect is relevant only to the RPC regime, it does not affect our upper limits on $\phot$. 

\subsection{Photoionization by shocked gas}
The fast winds detected in luminous quasars are expected to produce shocks. While the inner wind shocks most likely do not generally cool for nuclear winds with initial speeds $v_{\rm in} \gtrsim 10,000$ km s$^{-1}$ representative of BALs, the slower outer shocks with the ambient medium will often be radiative \citep[][]{FaucherGiguereQuataert12}.  
The observed narrow emission lines in quasars could in principle originate from gas ionized by this cooling radiation (\citealt{DopitaSutherland95}), rather than from gas ionized by the quasar radiation, as assumed above.
Curiously, the extended \oiii\ emission in quasars does not display the conical morphology expected by photoionization by the central source, but rather a spherical morphology (\citealt{Liu+13a}). The spherical morphology is perhaps more consistent with the shock scenario, since shocks can curve around obstacles (e.g.,\ \citealt{Wagner+13}), and therefore shock-powered emission can be produced even in regions not illuminated by the quasar. However, the apparent spherical morphology of ionized regions is subject to uncertainties in the point spread function of the observations so that more work is needed to put this finding on a firmer basis.

A concern with the shock scenario is whether the shock kinetic energy is sufficiently efficiently converted into [O~III] luminosity to explain the NLR emission (\citealt{Laor98}). 
In the most optimistic scenarios, the shock models of \cite{Allen+08} predict that $\zeta \approx10\%$ of the shock kinetic energy can be converted into [O~III] luminosity. 
Consider a luminous quasar with a fiducial wind kinetic luminosity $L_{\rm kin}=0.05L$, at the upper end of observationally-inferred values for galaxy-scale outflows powered by AGN, and assume that a fraction $f$ of the shock cooling radiation is actually re-processed by line-emitting clouds ($f$ thus plays the role of a covering factor). 
Then a rough upper limit on the expected [O~III] luminosity is $L_{\rm [O~III]}=0.005 f (\zeta/0.1)(L_{\rm kin}/0.05L) L$. 
In $L_{46}=1$ quasars, a typical bolometric conversion factor $L_{\rm [O~III]}\approx0.0003 L$ is inferred (\citealt{SternLaor12b}), indicating that a covering fraction $f>0.06$ might be sufficient for shocks to power the [O~III] emission in luminous quasars with powerful outflows. However this estimate rests on optimistic assumptions regarding the kinetic power of AGN winds and the efficiency $\zeta$, so more detailed modeling would be necessary to better quantify the theoretical viability of the shock excitation scenario for the NLR. 

Spectroscopic X-ray observations can empirically discriminate between the shock-ionization scenario and the quasar-ionization scenario. In the shock scenario, the X-ray spectrum should exhibit both line emission from the photoionized gas and free-free emission from the hot shocked gas (see fig.~2 in \citealt{Allen+08}). In contrast, in the quasar photoionization scenario, the emission lines are expected to be observed on top of a reflected hard X-ray quasar spectrum. High-resolution X-ray spectra are currently available only for low-luminosity AGN (\citealt{Bianchi+06, Wang+09, Wang+11, Wang+12}) but it would clearly be valuable to obtain such spectra for more luminous quasars. While low-luminosity AGN typically do not exhibit the free-free component expected in the shock scenario, recent observations suggest that only luminous AGN with $L>3\times10^{45}\ergs$ are capable of driving galaxy-wide outflows (\citealt{Veilleux+13, ZakamskaGreene14}) and so the NLR excitation mechanism could be different in those 
objects. 

\subsection{Magnetic fields}
In \S\ref{sec: pgas} we derived the thermal pressure of the ionized line-emitting gas under the assumption that it is not dominated by the magnetic pressure $\pmag$, i.e.\ 
\begin{equation}
 \pmag \equiv \frac{B_{\rm ion}^2}{8\pi} \lesssim \pgas ~,
\end{equation}
where the subscript `ion' emphasizes that we are referring to the magnetic field in the ionized gas. 
Since we find that $\pgas \sim \prad \equiv L/(4\pi r^2 c)$, the above inequality implies
\begin{equation}\label{eq: B field}
 B_{\rm ion} \lesssim 0.3\,L_{46}^{0.5} \left(\frac{r}{1\kpc}\right)^{-1} \mG ~.
\end{equation}
Do observed magnetic fields conform to this limit? Using measurements of Zeeman splitting in OH masers, \cite{McBride+14, McBride+15} find typical fields of $B_{\rm OH}\sim1\mG$ in the nuclei of ULIRGs. Taking this value as representative of quasar hosts, assuming the density of ionized gas is $10-100$ times lower than the density of OH masing clouds, and assuming that $B \propto n^{1/2} - n^1$, gives $B_{\rm ion} = 0.01 - 0.3 \mG$, consistent with equation~(\ref{eq: B field}). This suggests that our analysis is not significantly affected by magnetic pressure in the line-emitting clouds. If in reality $B_{\rm ion}$ is larger than estimated here, then the magnetic pressure in the clouds could contribute significantly to  balancing the external pressure (see eqn.~B1 in \citealt{Stern+14b}). Then, for a given external pressure the gas density in the cloud would be lower and hence the ionization level would be higher than derived above assuming $B_{\rm ion}=0$. In this case, our upper limits on $\phot$ are 
underestimated.

\subsection{Dust-less gas beyond the sublimation radius}
The lower left panel of Figure~\ref{fig: multiple line ratios} shows that there is not much room for narrow \Hb\ emission from dust-less gas, since it would imply that $\ldhb$ (the H$\beta$ luminosity from dusty gas) is only a fraction of the total observed $\Hb$ emission. Thus if the observed H$\beta$ emission were dominated by dust-less gas, the observed $\ldhbir$ ratio would be in tension with all the models considered. This supports our assumption that emission line gas beyond the sublimation radius is dusty. Further evidence that low-ionization gas beyond the sublimation radius is dusty is discussed in \S4 of S14. 
However, as discussed in S14, the composition of the high-ionization gas is less clear. The observed $\fevii/\nev$ ratio in the NLR is higher than expected from depleted abundances (\citealt{Shields+10} and references therein), suggesting the existence of highly ionized ($U\gtrsim0.1$) gas at $r>\rsub$ in which at least some of the dust is destroyed. 
To verify that the existence of such dust-less gas does not affect our conclusions, we compare the predictions of hydrostatic dust-less models with the observations shown in the top two panels of Figure~\ref{fig: multiple line ratios}. All other parameters are identical to the dusty models. 
Radiation pressure dominated dust-less models with $\phot/\prad=0.1$ predict $\lx/\loiii=0.02-1.2$ and $\nev/\neiii=0.2-1$, consistent with the observations. The maximum $\phot/\prad$ for which dust-less models are consistent with the observed $\lx/\loiii$ is $0.6$, while the maximum $\phot/\prad$ consistent with the observed $\nev/\neiii$ is $6.3$. Therefore, our main conclusions regarding the consistency of observed line ratios with RPC models are not qualitatively changed if dust-less clouds contribute to the NLR and eNLR emission, though the quantitative upper limits on $P_{\rm hot}/P_{\rm rad}$ are somewhat weakened. 

\subsection{Potential caveats in the analysis of the IR-emitting region}

To estimate the effect of unmodeled dust extinction of \Hb\ on the constraints derived in the IR panel of Figure~\ref{fig: multiple line ratios}, we compare the observed narrow line Balmer decrement to the predictions of our models. The observed mean NLR Balmer decrement in SDSS quasars in which both \Ha\ and \Hb\ are observed is 4.2 (\citealt{Shen+11}). We repeat our photoionization analysis with an external dust screen added to each photoionization model such that the predicted Balmer decrement matches the observed value of 4.2, and adjust the predicted $\ldhb$ according to extinction by this dust screen. We disregard models that predict a Balmer decrement larger than the observed value. 
The observed $\ldhbir$ shown in Figure~\ref{fig: multiple line ratios} is consistent with the range allowed by RPC models that include the effect of an external dust screen, and the new implied upper limit on $\phot$ is $< 0.6\prad$, somewhat higher than but not vastly different from the $\phot < 0.3\prad$ upper limit deduced from our fiducial models. 

Another potential caveat is that some of the IR-emitting clouds discussed in \S\ref{sec: IR} may be shielded by the smaller scale BLR (see fig.~1 in \citealt{Gaskell09}). 
Such `shielded' dusty clouds do not absorb any ionizing emission, since it is absorbed by the BLR, but do absorb the optical quasar emission to which the BLR clouds are transparent. Additionally, any energy transmitted by the BLR clouds, such as the optical emission lines, will also be absorbed by the shielded dusty clouds. Therefore, such clouds will contribute to $\lir$ but not to $\ldhb$. While it is plausible that such clouds have a significant contribution to the $3\mic$ emission, which originates from a distance $2-4$ times larger than the BLR (e.g., Fig.~13 in \citealt{Koshida+14}), we find it unlikely that such shielded clouds would dominate the $\lambda > 10\mic$ emission that originates at significantly larger scales, and contributes $\gtrsim50\%$ of $\lir$. Therefore, the true $\lir/\ldhb$ is unlikely to be more than twice the value shown in Figure~\ref{fig: multiple line ratios}. This uncertainty increases the upper limit derived on $\phot$ at IR-scales from $0.3$ to $0.6$.

\section{C. Free Adiabatic Expansion of shocked hot gas}\label{sec: free expansion}
We derive the dependence of $\phot$ on $r$ in the limit where the hot gas expands freely after shocking with the ambient ISM at a small radius $r_{\rm sw}$ (right panel of Fig.~\ref{fig: BAL_wind_structure} and \S\ref{sec: BAL wind}). We assume that as it expands, the wind cools adiabatically, and search for a spherically symmetric steady state solution. 
Mass conservation in a spherically symmetric geometry implies
\begin{equation}\label{eq: mass conservation}
 r^2\rho v = \const;
\end{equation}
adiabatic expansion implies
\begin{equation}\label{eq: adiabatic expansion}
 P^{1-\gamma}T^\gamma = \const;
\end{equation}
and momentum conservation implies
\begin{equation}\label{eq: momentum conservation}
\rho v \frac{\d v}{\d r} = - \frac{\d P}{\d r} - \frac{G M_t \rho}{r^2} ~.
\end{equation}
We can neglect gravity since the shocked wind has thermal velocities $\vnuc \sim 10,000 \kms\gg \sigma \sim 200 \kms$, the velocity dispersion of the galactic potential.
Looking for power-law solutions of the form 
$ \rho \propto r^l, T \propto r^m, v \propto r^n$, and assuming $\gamma=5/3$, we get
\begin{equation}
 \rho \propto r^{-3/2}~,~
 T    \propto r^{-1}~,~ 
 v    \propto r^{-1/2}~.
\end{equation}

\section{D. Notes on outflow momentum flux estimates}
\label{sec:momentum_appendix}
The data used to compute the outflow momentum fluxes compiled in Figure~\ref{fig: pdot} are taken from the references noted in the legend. For the \cite{Greene+14} measurement, we calculate the range of possible $\pdot$ from their preferred range of $\pdot v_{\rm out} = 10^{44}-10^{45}\ergs$ and assuming a velocity range $v_{\rm out}=250-1,000\kms$. 
For the UV absorption line measurements (blue points), the outflow energetics are estimated using the model of \cite{FaucherGiguere+12}. 
For the \cite{Bautista+10} measurement, we use their component `e'. 
To calculate $v_{\rm out}$ for the \cite{Liu+13b} objects, we assume following \cite{Harrison+14} that $v_{\rm out} = w_{80}/1.3$, where $w_{80}$ is the width of the part of the line that contains 80\% of the flux. 

\end{document}